\documentclass[aps,prd,reprint,nofootinbib,superscriptaddress]{revtex4-1}

\usepackage{ifxetex}
\ifxetex
   \usepackage{polyglossia}
   \setdefaultlanguage{english}
   \usepackage{fontspec}
\else
   \usepackage[utf8]{inputenc}
\fi

\usepackage{graphicx,color,overpic,mathtools}
\usepackage{amsthm}
\usepackage{bm}
\usepackage{amsmath}
\usepackage{amssymb}
\usepackage{setspace}
\usepackage{hyperref}
\PassOptionsToPackage{normalem}{ulem}
\usepackage{ulem}

\newcommand{\noprint}[1]{}

\newcommand{\tr}{{\text{tr}}}

\newcommand{\ii}{\mathrm{i}}
\renewcommand{\d}{\mathrm{d}}
\renewcommand{\Re}{\mathrm{Re}}
\renewcommand{\Im}{\mathrm{Im}}
\newcommand{\nn}{\nonumber}

\newtheorem{defi}{Definition}

\usepackage{braket}
\usepackage{bbm}

\providecommand{\abs}[1]{\lvert#1\rvert}
\providecommand{\norm}[1]{\lVert#1\rVert}

\begin{document}

\title{A non-perturbative analysis of spin-boson interactions using the Weyl relations}

\author{Jos\'e de Ram\'on}
\affiliation{Institute for Quantum Computing, University of Waterloo, Waterloo, Ontario, N2L 3G1, Canada}
\affiliation{Department of Applied Mathematics, University of Waterloo, Waterloo, Ontario, N2L 3G1, Canada}

\author{Eduardo Mart\'in-Mart\'inez}
\affiliation{Institute for Quantum Computing, University of Waterloo, Waterloo, Ontario, N2L 3G1, Canada}
\affiliation{Department of Applied Mathematics, University of Waterloo, Waterloo, Ontario, N2L 3G1, Canada}
\affiliation{Perimeter Institute for Theoretical Physics, 31 Caroline St N, Waterloo, Ontario, N2L 2Y5, Canada}

\begin{abstract}
We perform a non-perturbative analysis of the dynamics of a  two-level quantum system  subjected to repeated interactions with a bosonic environment when these interactions are intense and localized in time. We use the Weyl relations to obtain a closed expression for the resulting state of the spin despite the non-perturbative nature of the problem.  Furthermore, we study divisibility and memory effects in the dynamics and draw conclusions about the role that the quantum-mechanical features of the environment play on the dynamics of the two-level system. 
\end{abstract}
\maketitle

\section{introduction}

The theory of open quantum systems(OQS) studies the dynamics of quantum systems in the presence of an environment \cite{rivas2012open}. This theory is of great interest both from the experimental perspective, in which interactions with the environment do affect the outcomes of a particular experiment, and from pure theoretical purposes. Indeed, from the theoretical point of view, the mathematical tools used to describe open quantum systems extend to a much wider set of physical situations. By Naimark's dilation theorem \cite{paulsen_2003}, any transformation in the state of a quantum system in the form a linear, completely positive, and trace preserving (CPTP) map can be interpreted as an interaction with a reservoir in an enlarged system\footnote{The maps that generate dynamics for quantum systems through linear CPTP maps are commonly known in the literature of open quantum systems as universal dynamical maps (UDM) \cite{rivas2012open}}. Furthermore, the open quantum systems formalism can be used for tackling, at least partially, the problem of measurement in quantum theory, since a measurement process can be thought as an interaction with an instrument or probe \cite{von2013mathematische}.

With this generality in mind, the theory of open quantum systems has been regarded as the quantum analogue of the theory of stochastic processes \cite{cox2017theory}, which studies the dynamics of classical random variables. A central concept in the study of classical stochastic processes is the so-called Markovian property, which is related with memory effects in the dynamics and plays a key role from both technical and conceptual points of view. In a classical set-up, Markovianity and memory are explicitly related \cite{cox2017theory}. However, it is still an open problem to define an analogue of the Markovian property---and its relation with  memory effects in the dynamics---in the quantum case. 

Markovian processes are often referred to as ``memoryless'': the system's dynamics depends only on the immediate past of the system and not on its whole past history. However, memory effects can play a significant role in many physical proceses, thus justifying the study of non-Markovian dynamics. Indeed, memory effects have been suggested  to be useful in quantum information processing \cite{Dong_2018}, quantum optics \cite{Dinc2019exactmarkoviannon}, and quantum thermodynamics \cite{Styliaris:2019aa}. In this paper we will follow the definition of Markovianity suggested in \cite{Rivas_2014}, which we briefly summarize in section \ref{sec:Marcopolo}. 

We will analyze non-Markovian effects in a particular scenario. Namely, the interaction of a two-level quantum system locally coupled to a bosonic environment. We analyze the behavior of the two-level system when it interacts repeatedly with its environment through fast and intense couplings. We model this situation with an interaction Hamiltonian representing a spin-boson coupling which is repeatedly switched on and off, i.e. modulated with a train of Dirac delta functions. This will allow us to obtain analytic, non-perturbative results. 

In this work we will avoid to resort to approximations that are common in many standard analyses in quantum optics and atomic physics. Examples of such approximations are  the rotating wave approximation and the single (or few) mode approximation \cite{Scully_1997}. It has been shown that these approximations fail in many scenarios \cite{Lopp_2018,Funai:2019aa,Compagno_1990,Compagno_1989,Clerk_1998}, and they are incompatible with our approach from a technical point of view, as it will be clear throughout the paper. Our coupling model is inspired by the Unruh-DeWitt model (UDW) \cite{Hawking:1979ig}, which can be shown to capture the main features of light-matter interaction when exchange of angular momentum is irrelevant \cite{PhysRevD.94.064074}, or the phenomenology of superconducting circuits coupled to transmission lines \cite{PhysRevA.96.052325}. Moreover, the UDW model is been shown  appropriate to describe light-matter interactions in scenarios where relativistic effects are relevant \cite{PhysRevD.97.105026}, therefore the results presented here are relevant to the study the role of memory effects in relativistic set-ups.

Within the UDW model literature, Dirac delta switching functions have been extensively used in the recent years to extract non-perturbative results \cite{Hotta:2008aa,PhysRevD.96.065008,PhysRevD.98.105011}. In this paper we present an elementary proof of results involving them. Remarkably, we will show that this scenario is equivalent to a collision model \cite{Lorenzo_2017,Ciccarello_2013,PhysRevLett.123.180602,Ciccarello:2013aa,PhysRevA.100.042702,Filippov:2017aa,Ryb_r_2012} in which correlations in the environment are considered, both of classical and quantum origin. These collision models with correlations have been already used to study non-Markovian dynamics, e.g. for qubits interacting with a fermionic environment, see  \cite{Filippov:2017aa}.

 Collision models are usually intended to simulate continuum dynamics in some limit, for instance, when the number of collisions per unit of time is large. For most of this paper, however, we will consider insightful enough to focus in the case of two interactions. It has been argued \cite{Ryb_r_2012} that some of the main features of the dynamics can be captured in this way, and this perhaps oversimplified analysis will allow us to display our analytical results in full generality. We will further consider cases in which the interactions with the environment are synchronized with the internal dynamics of the two-level system, showing that these generate pure-dephasing dynamics \cite{MR1796805}. 

We present and motivate the physical system of interest in section \ref{deltatocollision}, and derive the dynamics of the two-level system induced by its interaction with the bosonic reservoir in the form of a CPTP map. In section \ref{environment} we describe the bosonic environment together with its algebraic properties. More concretely, we introduce a central algebraic property of bosonic systems, the 
Weyl relations, and describe properties of the main class of bosonic states considered in the work,  Gaussian states.

Section \ref{single} is devoted to study the case of a single interaction, whose analysis and interpretation will be essential in the following sections. In section \ref{muchas} we derive the general form of the CPTP map with an arbitrary number of interactions, which we solve exactly in section \ref{double} for the case of two interactions and in section \ref{dephasing} for an arbitrary number of synchronized interactions. 

The main section of this paper is section \ref{sec:Marcopolo}, in which we study memory effects in the cases previously presented. We supplement this section with a short introduction to memory effects in quantum mechanics to contextualize our work and to assist the unfamiliar reader. 

Finally, we conclude, summarize, and discuss future work in section \ref{conclusion}.

\section{From spin-environment interactions to complete positive maps} \label{deltatocollision}

We consider a two-level quantum system subject to repeated interactions with an environment. Namely, the interaction of the system and the environment will be switched on and off for a finite number of times.  Consider that the evolution of  the system and the environment is described by the rather general Hamiltonian
\begin{align}
\hat{H}&=\hat{H}_{\textsc{s}}+\hat{H}_{\textsc{E}}+\hat{H}_{\textsc{se}},
\end{align}
where  $\hat{H}_{\textsc{E}}$ denotes the free evolution of the environment, which we leave arbitrary.  $\hat{H}_{\textsc{s}}=\Omega \mathbf{h}\cdot \bm{\hat{\sigma}}$ is the free Hamiltonian of the two-level system. The spin-environment interaction Hamiltonian is given by
\begin{equation}\label{hamiltonian}
    \hat{H}_{\textsc{se}}= \chi(t) \bm{\alpha}\cdot \bm{\hat{\sigma}} \otimes \hat{{O}}.
\end{equation}
Here  $\mathbf{h}\cdot \bm{\hat{\sigma}}$ and $\bm{\alpha}\cdot \bm{\hat{\sigma}}$ are Pauli observables in the directions $\mathbf{h}$ and $\bm{\alpha}$ respectively.  $\hat{{O}}$ is an observable of the environment, that is, a self-adjoint operator that also commutes with all observables of the spin.

Finally $\chi(t)$ is a switching function that regulates the rate and intensity of the interactions between the two-level system and the environment. Note that outside the support of $\chi$ the two subsystems evolve independently.

If the system starts at $t=0$ in a state described by the density matrix $\hat{\rho}^0$, then the state of the system at a later time $t$ is given by the unitary evolution
\begin{align}
    \hat{\rho}(t)=\hat{U}(t)\hat{\rho}^0\hat{U}^{\dagger}(t),
\end{align}
where the unitary operator $\hat{U}$ is given by
\begin{align}\label{Dyson1}
\nn\hat{U}(t)=&\mathcal{T}e^{-\ii\int_{0}^{t} \d t \hat{H}(t)}\\
&=\sum_{n=0}^{\infty}\frac{(-\ii)^n}{n!}\int_{0}^{t}...\int_0^{t_1}\d^n t\;  \hat{H}(t_n)...\hat{H}(t_0).
\end{align}
It is useful to define the evolution in the interaction picture:
\begin{align}
    \hat{\rho}_{I}(t)=\hat{U}_{I}(t)\hat{\rho}^0\hat{U}_{I}^{\dagger}(t),
\end{align}
where this time the unitary evolution is given by  
\begin{equation}\label{Uinteraction1}
\hat{U}_I(t)=\mathcal{T}e^{-\ii\int_0^{t} \d t\chi(t) \hat{H}_{I}(t)},
\end{equation}
that is defined analogously to \eqref{Dyson1}, and where $\hat{H}_{I}(t)$ is the interaction Hamiltonian in the interaction picture. For the Hamiltonian \eqref{hamiltonian}, this acquires the form
\begin{align}
     \hat{H}_{I}(t)=\hat{U}_{\text{free}}^{\dagger}(t)\hat{H}_{\textsc{se}}(t)\hat{U}_{\text{free}}(t)=\bm{r}(t)\cdot\bm{\hat{\sigma}}\otimes\hat{O}(t)
\end{align}
where
\begin{align}
    \hat{O}(t)=\left(\mathcal{T}e^{-\ii\int_{0}^{t} \d t \hat{H}_{\textsc{e}}(t)}\right)^{\dagger}\hat{O}\;\mathcal{T}e^{-\ii\int_{0}^{t} \d t \hat{H}_{\textsc{e}}(t)}
\end{align}
and
\begin{align}\label{erredete}
    \nn&\bm{r}(t)=(\bm{h}\cdot{\bm{\alpha}})\bm{h}+\\
    &\cos(\Omega t) \left(\bm{\alpha} -(\bm{h}\cdot\bm{\alpha}\right)\bm{h})-\sin(\Omega t)(\bm{h}\times\bm{\alpha}).
\end{align}
Finally, the Schrodinger picture is recovered by just applying the free evolution to the state in the interaction picture, that is 
\begin{align}
    \hat{\rho}(t)=\hat{U}_{\text{free}}(t)\hat{\rho}_{I}\hat{U}_{\text{free}}^{\dagger}(t).
\end{align}

We recall that the duration and the intensity of the interaction are regulated by the ``switching function'' $\chi(t)$. We assume that the function has compact support and we also assume that the interaction starts at some time $t\ge 0$. In other words, we say that there exists a $t_c$ such that
\begin{align}
    \text{supp}\{\chi(t)\}\subset [0,t_c].
\end{align}

Therefore, for all $t>t_c$, the integrals in \eqref{Uinteraction1} can be extented from their domain to the whole real line without changing their value.  Then it can be shown that the evolution in the interaction picture takes the form
\begin{align}\label{Uinteraction2}
\nn &\hat{U} _I(t>t_c)=\mathcal{T}e^{-\ii\int\d t\hat{H}_{I}(t)}\\
&=\sum_{n=0}^{\infty}\frac{(-\ii)^n}{n!}\int_{\mathbb{R}^n}\!\!\!\d^n t\; \mathcal{T}\left[\hat{H}_{I}(t_1)...\hat{H}_{I}(t_n)\right],   
\end{align}
where $\mathcal{T}[\cdot]$ denotes the time-ordered product \cite{joachain1983quantum} and \mbox{$\d^n t=\d t_1\dots \d t_n$}. Henceforth, we will always assume that the time $t$ at which we evaluate the state is larger than the time the interaction ends: $t>t_c$.

Now we assume that we only have access to information from the two-level system. To compute the partial state, $\hat{\rho}_{\textsc{q}}(t)$, we trace over the environment degrees of freedom:
\begin{align}
   \nn  \hat{\rho}_{\textsc{q}}(t)&=\text{tr}_{E}\left[ \hat{\rho}(t)\right]=e^{-\ii\Omega \mathbf{h}\cdot \bm{\hat{\sigma}}t}\text{tr}_{E}\left[ \hat{\rho}_I(t)\right]e^{\ii\Omega \mathbf{h}\cdot \bm{\hat{\sigma}}t}\\
   &= e^{-\ii\Omega \mathbf{h}\cdot \bm{\hat{\sigma}}t}\text{tr}_{E}\left[ \hat{U}_{I}\hat{\rho}^0\hat{U}_{I}^{\dagger}\right]e^{\ii\Omega \mathbf{h}\cdot \bm{\hat{\sigma}}t}.
\end{align}

Finally, we assume that at $t=0$ the spin and the environment are completely uncorrelated, i.e., the initial state has the form of the tensor product
\begin{equation}
    \hat{\rho}^0=\hat{\rho}_{\textsc{q}}^0\otimes\hat{\rho}_{\textsc{e}}.
\end{equation}
This last assumption allows one to write the evolution in the interaction picture in terms of a completely positive trace-preserving (CPTP) map, given by the quantum channel
\begin{align}
 \hat{\rho}_{\textsc{q}}(t)= e^{-\ii\Omega \mathbf{h}\cdot \bm{\hat{\sigma}}t} \mathcal{E}[\hat{\rho}_{\textsc{q}}^0]e^{\ii\Omega \mathbf{h}\cdot \bm{\hat{\sigma}}t}.
\end{align}
where
\begin{equation}\label{CPTP}
    \mathcal{E}[\hat{\rho}_{\textsc{q}}^0]=\text{tr}_{E}\left[ \hat{U}_{I}\hat{\rho}_{\textsc{q}}^0\otimes\hat{\rho}_{\textsc{e}}\hat{U}_{I}^{\dagger}\right].
\end{equation}
Note that, since we are considering times after the interaction is finished $t>t_c$, the channel $\mathcal{E}$ does not explicitly depend on $t$.

In the following we will consider a special case of switching function $\chi(t)$ describing successions of very fast interactions of short duration being repeatedly switched on and off throughout the interval $[0,t_c]$. Concretely, we focus on the limit when the interaction times of each instance of switching is much shorter than any other timescale in the system. Namely, we consider a scenario in which the interaction is different from zero only in a discrete set of points $t_1...t_n$.  In order to describe this situation the switching function has to be different from zero only in this discrete set of points. In addition, its value at these points has to be defined carefully in order to generate evolution. As the duration of the interaction is taken to zero, the strength of the interaction increases. In other words, we consider the switching function to come from a limit of nascent deltas.  Therefore, it seems sensible to consider a switching function of the form  
\begin{align}\label{deltas}
    \chi(t)&=\sum_{k=0}^N \delta\left({t-{t}_k}\right) & {t}_N>&...>{t}_0.
    \end{align}
    
As shown in appendix \ref{deltasproof}, for such a  switching function, the unitary evolution of the joint system factorizes, i.e., it can be written as the time-ordered product
\begin{align}\label{ues}
  \hat{U}_{I}= e^{-\ii \hat{H}_{I}( {t}_N)}...e^{-\ii \hat{H}_I( {t}_0)}.
\end{align}
where ${t}_N>...>{t}_0$.

Although this assumption of delta-coupling interactions could seem at first perhaps naive, we will show in the section \ref{muchas} that it actually accounts for a number of widely accepted models in open quantum systems, commonly know as collision models \cite{PhysRevLett.123.180602,PhysRevA.100.042702,Ciccarello_2013,Ciccarello:2013aa,Filippov:2017aa,Lorenzo_2017,PhysRevA.97.052120,PhysRevA.94.032126,PhysRevA.95.042114}. 

\section{Analysis of the environment:\\
Weyl relations}\label{environment}

For this paper, we consider  that the operator $\hat{O}$ is a fundamental observable of a representation of the canonical commutation relations (CCR) \cite{MR0493420}. For example, the Hamiltonian \eqref{hamiltonian} could correspond to a linear coupling of a spin to a quadrature of the oscillator (e.g., $\hat O=\hat X$, $\hat X$ being the position operator of a harmonic oscillator), it could represent a linear coupling in the context of general spin-boson models, for example between a two level system and a set of modes of an optical field, as in the exact Rabi or Glauber models in quantum optics (without rotating wave approximation \cite{Scully_1997}), or it could represent the common models of coupling particle detectors to quantum fields in the context of QFT in curved spacetimes, where then $\hat O$ can be either the field amplitude or its canonical conjugate momentum (or any combination of the two) \cite{PhysRevD.97.105026}.

For CCR representations we can build a Hilbert space $\mathcal{H}_1$, commonly known as the ``one particle'' Hilbert space. This Hilbert space can be understood as the space of classical solutions of some linear differential equations, and the fundamental observables $\{\hat{O}_i\}$ are then operators that generate the one-particle Hilbert space by acting over a cyclic vector usually called vacuum state \cite{wald1994quantum}. 

The full Hilbert space is then constructed by linear combinations of symmetrized tensor products of the one-particle Hilbert space, thereby building a so-called Fock representations of the CCR. It can be shown that for these fundamental observables it holds that  
\begin{align}\label{commutator} [\hat{O}_i,\hat{O}_j]=\mathcal{C}_{ij}\openone_{\mathcal{H}_{\textsc{e}}},
\end{align}
where $\mathcal{H}_{\textsc{e}}$ is given in this case by the Fock space, and $\mathcal{C}_{ij}$ is a constant. If the Hamiltonian is quadratic, then it implements a canonical transformation, i.e., evolution takes fundamental observables into fundamental observables. Then, given a fundamental observable  $\hat{O}$, its unitary evolution defines a one-parameter family, of fundamental observables $\hat{O}$ that fulfills
\begin{align}
    [\hat{O}(t),\hat{O}(t')]\propto \mathcal{C}(t,t') \openone_{\mathcal{H}_{\textsc{e}}},
\end{align}
where $\mathcal{C}$ is a scalar, antisymmetric function of $t,t'$.

This means that the expectation value of \eqref{commutator} is henceforth state independent. Also, note that since $\hat{O}(t)$ is a self-adjoint operator then $\mathcal{C}(t,t')$ is pure imaginary. Indeed,
\begin{align}
 \nonumber & \mathcal{C}(t,t')^*\openone_{\mathcal{H}_{\textsc{e}}}=  [\hat{O}(t),\hat{O}(t')]^{\dagger}=[\hat{O}^{\dagger}(t'),\hat{O}^{\dagger}(t)]\\
   &=[\hat{O}(t'),\hat{O}(t)]=- [\hat{O}(t),\hat{O}(t')]=-\mathcal{C}(t,t')\openone_{\mathcal{H}_{\textsc{e}}}.
\end{align}

The commutation relations \eqref{commutator} can be represented in terms of the exponentials of the fundamental observables, that is
\begin{align}
    e^{-\ii\hat{O}_i}e^{-\ii\hat{O}_j}=e^{-\mathcal{C}_{ij}}e^{-\ii\hat{O}_j}e^{-\ii\hat{O}_i},
\end{align}
that has as a consequence
\begin{align}\label{Weyl}
    e^{-\ii\hat{O}_i}e^{-\ii\hat{O}_j}=e^{-\frac{1}{2}\mathcal{C}_{ij}}e^{-\ii\hat{O}_i-\ii\hat{O}_j}.
\end{align}
This exponential version of the commutation relations is also known as Weyl relations \cite{MR0493420,wald1994quantum}. Technically, both representations are not equivalent in general. The Weyl relations \eqref{Weyl} always imply commutation relations of the form \eqref{commutator} as a consequence of Stone's theorem \cite{MR751959}. The converse, however, is not true. This has to do with the unbounded character of any algebra of operators satisfying the relations \eqref{commutator}, and the domain in which these relations hold.

Handling commutation relations of unbounded operators is mathematically subtle \cite{MR0493420}. There are well-known examples where the exponentials of two self-adjoint operators do not commute even when the operators that generate them do in their common domain. This is illustrated by e.g., Nelson's example \cite{MR0493420}. For this reason, from a fundamental level, it makes more sense to define the operators in a CCR through the exponential version of the commutation relationships \eqref{Weyl} rather than \eqref{commutator}. The imaginary exponential of  unbounded self-adjoint operators become unitary, hence bounded operators, removing the subtleties we mentioned above. For this work we will take equation \eqref{Weyl} as fundamental. This is the approach taken, for example, in some formulations of algebraic quantum field theory and other areas of mathematical physics \cite{Fewster:2019aa}.

In addition, in this work we will eventually consider that the environment's state $\hat{\rho}_{\textsc{e}}$ is quasi-free, also known as Gaussian state \cite{1751-8121-51-24-245301,wald1994quantum,Fewster:2019aa}. A Gaussian state is defined for CCR representations in such a way that, for any $\hat{O}$ belonging to the CCR representation, we have
\begin{align}
    &\big\langle{e^{-\ii\hat{O}}}\big\rangle=e^{-\ii\braket{\hat O}-\frac{1}{2}\left(\braket{\hat{O}^2}-\braket{\hat O}^2\right)}=e^{-\ii\braket{\hat O}}e^{-\frac{1}{2}\big\langle{\left(\hat{O}-\braket{\hat O}\right)^2}\big \rangle},
\end{align}
where $\langle\hat O\rangle_{\hat \rho}=\tr{[\hat O\hat \rho]}$. A particularly simple kind of Gaussian states are those $\braket{\hat O}=0$, called even Gaussian states, for which 
\begin{align}
    \big\langle e^{-\ii\hat{O}}\big\rangle=e^{-\frac{1}{2}\braket{\hat{O}^2}}.
\end{align}
Not-even Gaussian states describe the so-called coherent states of the theory. Notice, however, that in quantum optics the term coherent is reserved for pure non-even Gaussian states. 

Thermal states of the environment are always even Gaussian states \cite{wald1994quantum,Fewster:2019aa}. For this reason, in this work we will mainly focus on even Gaussian states. Notice in any case that given a state of the environment $\hat{\rho}_{\textsc{e}}$ we can always define a translated operator $\hat{O}_i'=\hat{O}_i-\braket{\hat O_i}\openone$ for which the Gaussian state is even. The result of this shift of the operator $\hat{O}$ in the spin-boson interaction discussed previously can be seen, roughly speaking, as a decoupling of the interaction of the spin with the mean value of $\hat O$ (the observable in the environment), but is still coupled to its fluctuations.

For the rest of this article it will be useful to consider the following results. First, is is easy to check that given the product of a sequence of $N$ elements of the exponential algebra, it fulfills
\begin{align}
    e^{-\ii\hat{O}_N}...e^{-\ii\hat{O}_1}=e^{-\ii\sum_{i=1}^N\hat{O}_i}e^{-\frac{1}{2}\sum_{i=1}^{N}\sum_{j=1}^{i-1}\mathcal{C}_{ij}},
\end{align}
which comes from applying \eqref{Weyl} recursively from left to right.
Further, its expectation  value in a general Gaussian state $\hat{\rho}_{\textsc{e}}$ is 
\begin{align}\label{mierdasgaussianas2}
  \nn  &\braket{e^{-\ii\hat{O}_N}...e^{-\ii\hat{O}_0}}_{\hat{\rho}_{\textsc{e}}}\\
  \nn  &=\braket{e^{-\ii\sum_{i=1}^N\hat{O}_i}}_{\hat{\rho}_{\textsc{e}}}e^{-\frac{1}{2}\sum_{i=0}^{N}\sum_{j=1}^{i-1}\mathcal{C}_{ij}}\\
 \nn& =e^{-\ii\sum_{i=1}^{N}\braket{\hat{O}_i}_{\hat{\rho}_{\textsc{e}}}}e^{-\frac{1}{2}\sum_{i=1}^{N}\braket{(\hat{O}_i-\braket{\hat{O}_i}_{\hat{\rho}_{\textsc{e}}}{\openone})^2}_{\hat{\rho}_{\textsc{e}}}}\\
    \nn&\times e^{-\frac{1}{2}\sum_{i=1}^{N}\sum_{j=1}^{i-1}\braket{\{\hat{O}_i-\braket{\hat{O}_i}_{\hat{\rho}_{\textsc{e}}}\openone,\hat{O}_j-\braket{\hat{O}_i}_{\hat{\rho}_{\textsc{e}}}\openone\}}_{\hat{\rho}_{\textsc{e}}}}\\
    \nn&\times e^{-\frac{1}{2}\sum_{i=1}^{N}\sum_{j=1}^{i-1}\mathcal{C}_{ij}}\\
    \nn   &=e^{-\ii\sum_{i=1}^{N}\braket{\hat{O}_i}_{\hat{\rho}_{\textsc{e}}}}e^{-\frac{1}{2}\sum_{i=1}^{N}\braket{(\hat{O}_i-\braket{\hat{O}_i}_{\hat{\rho}_{\textsc{e}}}\openone)^2}_{\hat{\rho}_{\textsc{e}}}}\\
    &\times e^{-\sum_{i=1}^{N}\sum_{j=1}^{i-1}\braket{(\hat{O}_i-\braket{\hat{O}_i}_{\hat{\rho}_{\textsc{e}}})(\hat{O}_j-\braket{\hat{O}_j}_{\hat{\rho}_{\textsc{e}}}\openone)}_{\hat{\rho}_{\textsc{e}}}},
\end{align}
where the anti-commutator is \mbox{$\{\hat{O}_i,\hat{O}_j\}=\hat{O}_i\hat{O}_j+\hat{O}_j\hat{O}_i$}.
Note that for even Gaussian states the expression takes a particularly simple form:
\begin{align}\label{mierdasgaussianas}
  \nn  &\braket{e^{-\ii\hat{O}_N}...e^{-\ii\hat{O}_0}}_{\hat{\rho}_{\textsc{e}}}\\
 \nn& =e^{-\frac{1}{2}\braket{(\sum_{i=1}^N\hat{O}_i)^2}_{\hat{\rho}_{\textsc{e}}}}e^{-\frac{1}{2}\sum_{i=1}^{N}\sum_{j=1}^{i-1}\mathcal{C}_{ij}}\\
&=e^{-\frac{1}{2}\sum_{i=1}^{N}\braket{\hat{O}_i^2}_{\hat{\rho}_{\textsc{e}}}}e^{-\sum_{i=1}^{N}\sum_{j=1}^{i-1}\braket{\hat{O}_i\hat{O}_j}_{\hat{\rho}_{\textsc{e}}}},
\end{align}

Finally, note that,  the real and imaginary parts  $\braket{\hat{O}_i\hat{O}_j}_{\hat{\rho}_{\textsc{e}}}$ are related to the commutator and the anti-commutator
\begin{align}\label{real}
    \Re\braket{\hat{O}_i\hat{O}_j}_{\hat{\rho}_{\textsc{e}}}=\frac{\braket{\{\hat{O}_i,\hat{O}_j\}}_{\hat{\rho}_{\textsc{e}}}}{2},\quad \Im\braket{\hat{O}_i\hat{O}_j}_{\hat{\rho}_{\textsc{e}}}=\frac{-\ii\mathcal{C}_{ij}}{2}.
\end{align}

\section{Single interaction}\label{single}

In this section we analyze the interaction of the spin with an environment when the interaction takes place in a single instant. In the case of a single interaction at $t_0$, the unitary evolution of the system plus environment is 
\begin{align}\label{uniuno}
    \hat{U}_{t_0}=e^{-\ii\bm{r}({t}_0)\cdot\bm{\hat{\sigma}}\otimes\hat{O}({t}_0)},
\end{align}
 which follows from a straightforward particularization of equation \eqref{ues} with one interaction and with the Hamiltonian given by \eqref{hamiltonian}.

 The operator \eqref{uniuno} admits a partial spectral decomposition in the eigenprojectors of the Pauli operator $\bm{r}({t}_0)\cdot\bm{\hat{\sigma}}$, i.e.,
\begin{align}
   \hat{ U}_{t_0}=\hat{P}^+(t_0)\otimes e^{-\ii\hat{O}({t}_0)}+\hat{P}^-(t_0)\otimes e^{\ii\hat{O}({t}_0)},
\end{align}
where
\begin{align}
    \hat{P}^{+,-}=\frac{\openone\pm\bm{r}({t_0})\!\cdot\!\bm{\hat{\sigma}}}{2}.
\end{align}
are the orthogonal projectors associated with the eigenvalues $\pm 1$ of the Pauli observable.

After applying the unitary operator \eqref{uniuno} to the spin plus environment, the resulting channel \eqref{CPTP} over the reduced state is then 
\begin{align}\label{CPTP1d}
   \nn\mathcal{E}_{t_0}[\hat{\rho}_{\textsc{q}}^0]&=\hat{P}^-(t_0)\hat{\rho}_{\textsc{q}}^0\hat{P}^-(t_0) +  \hat{P}^+(t_0)\hat{\rho}_{\textsc{q}}^0\hat{P}^+(t_0)\\
   & + \gamma_{+-}\hat{P}^+(t_0)\hat{\rho}_{\textsc{q}}^0\hat{P}^-(t_0) + \gamma_{-+} \hat{P}^-(t_0)\hat{\rho}_{\textsc{q}}^0\hat{P}^+(t_0),
\end{align}
where the factors $\gamma_{+-}$ are given by
\begin{align}
    \gamma_{+-}=\gamma_{-+}^*=\braket{e^{-2\ii\hat{O}({t}_0)}}_{\hat{\rho}_{\textsc{e}}},
\end{align}
 For future convenience, the channel \eqref{CPTP1d} can be written equivalently as
\begin{align}
  \nn   \mathcal{E}_{t_0}[\hat{\rho}_{\textsc{q}}^0]&=\frac{1}{2}(\text{Id}+\mathcal{U}(t_0))[\hat{\rho}_{\textsc{q}}^0]\\
  &\nn + \frac{\Re \braket{e^{-2\ii\hat{O}({t}_0)}}_{\hat{\rho}_{\textsc{e}}}}{2}(\text{Id}-\mathcal{U}(t_0))[\hat{\rho}_{\textsc{q}}^0]\\
  &+\frac{\ii\Im \braket{e^{-2\ii\hat{O}({t}_0)}}_{\hat{\rho}_{\textsc{e}}}}{2}\left[\bm{r}({t})\!\cdot\!\bm{\hat{\sigma}},\hat{\rho}\right].
\end{align}
where we Id denotes the identity channel and
\begin{align}
    \mathcal{U}(t)[\hat \rho]=\bm{r}({t})\!\cdot\!\bm{\hat{\sigma}}\hat{\rho}\;\bm{r}({t})\!\cdot\!\bm{\hat{\sigma}}
\end{align}
is a unitary channel.
 
 Therefore, the result of the interaction is a phase damping in the eigenbasis of $\bm{r}({t})\!\cdot\!\bm{\hat{\sigma}}$ plus a rotation  of axis $\bm{r}$. Indeed, since $0<|\gamma_{+-}|<1$, we can parametrize the factor as $\gamma_{+-}=e^{-\alpha}e^{\ii\phi}$, with $\alpha>0$. Then if the input state is written  in the eigenbasis of $\bm{r}({t})\!\cdot\!\bm{\hat{\sigma}}$ as
 \begin{align}
     \hat\rho=
     \begin{pmatrix}
     \rho_{++}&\rho_{+-}\\
    \rho_{-+} &\rho_{--}
     \end{pmatrix},
 \end{align}
  the channel gives an output
  \begin{align}
  &\nn   \hat\rho=
     \begin{pmatrix}
     \rho_{++}&e^{-\alpha}e^{\ii\phi}\rho_{+-}\\
    e^{-\alpha}e^{-\ii\phi}\rho_{-+} &\rho_{--}
     \end{pmatrix}=\\
     &e^{\frac{\ii}{2}\phi\bm{r}({t})\cdot\bm{\hat{\sigma}}}
     \begin{pmatrix}
     \rho_{++}&e^{-\alpha}\rho_{+-}\\
    e^{-\alpha}\rho_{-+} &\rho_{--}
     \end{pmatrix}
     e^{\frac{-\ii}{2}\phi\bm{r}({t})\cdot\bm{\hat{\sigma}}}.
 \end{align}
 
 Note that the phase of $\gamma_{+-}$ only implements a unitary transformation. 
 
 It is particularly interesting to consider the action of this channel when the environment is in a general Gaussian state, for which the factor $\gamma_{+-}$ takes the form
 \begin{align}
   \gamma_{+-}=e^{-2\ii\braket{\hat O(t_0)}_{\hat{\rho}_{\textsc{e}}}}e^{-2\braket{\left(\hat{O}(t_0)-\braket{\hat O(t_0)}_{\hat{\rho}_{\textsc{e}}}\openone\right)^2}_{\hat{\rho}_{\textsc{e}}}},
\end{align}
we see that, on one hand, the first moment of $\hat{O}(t_0)$ only implements a unitary transformation and does not play a relevant role in the damping of the phase, and thus in the entanglement induced by the interaction between the spin and the environment, as it was already observed, for example, in \cite{PhysRevD.98.085007}. On the other hand, the fluctuations of the environment encoded in  $\big\langle(\hat{O}(t_0)-\braket{\hat O(t_0)}_{\hat{\rho}_{\textsc{e}}})^2\big\rangle_{\hat{\rho}_{\textsc{e}}}$ are the true  responsible for the phase damping.

 Since in this simple case the interaction takes place in a single instant, the evolution of the environment after the interaction is irrelevant for the spin's dynamics. With just one interaction of such kind, even when entanglement is generated between the spin and the environment, the effective dynamics of the spin can be thought as coming from an external random force, rather than from a dynamical process in which the environment back-reacts. Note that, interestingly enough, this feature remains true regardless of the quantum nature of the environment. Note further that since with one interaction the relevant influence of the environment on the spin is through $\braket{e^{-2\ii\hat{O}({t}_0)}}_{\hat{\rho}_{\textsc{e}}}$, it is impossible to distinguish the non-commutative character of the environment observables, since $\braket{e^{-2\ii\hat{O}({t}_0)}}_{\hat{\rho}_{\textsc{e}}}$ can be interpreted as the characteristic function of some random variable, and thereby the quantum character of the environment stands irrelevant in regards to the effective action on the spin.

We proceed now to further analyze the dynamics of the channel on the spin after a strong-point, delta-like interaction in time with the environment. For simplicity, we constrain ourselves to even Gaussian states of the environment. Since, for this case, $\gamma_{+-}$ is real, the effect of the interaction is just a phase damping in the eigenbasis of $\bm{r}({t})\cdot\bm{\hat{\sigma}}$. Of course, depending on the relative orientation of $\bm{r}({t_0})$ with respect to $\bm{h}$, the channel can be interpreted as a bit flip, a phase flip or a bit-phase flip \cite{MR1796805}. 

For our purposes it will be useful to write the channel \eqref{CPTP1d} as a function from $\mathbb{R}^3\to\mathbb R^3$, i.e., as a function of the Bloch vector of the initial state $\hat{\rho}_{\textsc{q}}^0$. If we write the state $\hat{\rho}^0_{\textsc{q}}$ in terms of its Bloch vector $\bm{u}$, that is,
\begin{align}\label{Bloch}
    \hat{\rho}^0_{\textsc{q}}=\frac{\openone+\bm{u}\cdot\bm{\hat{\sigma}} }{2} 
\end{align}
(where $u_i=\tr[\hat{\rho}^0_{\textsc{q}}\hat\sigma_i]$),
\begin{align}
    \mathcal{E}_{t_0}[\hat{\rho}_{\textsc{q}}^0]=\frac{\openone+\bm{\mathcal{E}}_{t_0}(\bm u)\cdot\bm{\hat{\sigma}} }{2},
\end{align}
where
\begin{align}\label{CPTP1dBloch}
    \bm{\mathcal{E}}_{t_0}(\bm u)=
    \gamma_{+-}\bm u +(1-\gamma_{+-})(\bm u\cdot \bm r(t_0))\bm r_0. 
\end{align}

Note that the channel is unital, and thus it cannot decrease the entropy of the system \cite{Lindblad_1973}. It is known that for finite dimensional systems a necessary and sufficient condition for a completely positive trace preserving channel $\mathcal{E}$ to be able to decrease the entropy of a state is to be non-unital, that is
\begin{align}
    \mathcal{E}[I]\neq I.
\end{align}
It is trivial to check that it is sufficient, as a non-unital channel always changes the maximally mixed state thus decreasing the entropy of the system, for the maximally mixed state maximizes the entropy.
The necessity is not but a consequence of the monotony of the relative entropy under CPTP channels \cite{Lindblad_1973} (See e.g., Appendix \ref{entropygain} for a quick proof).  

We can evaluate the purity after the strong-point interaction with the environment, that is
\begin{equation}
    \mathcal{P}(\hat{\rho}_{\textsc{q}})=\text{tr}[\hat{\rho}^2_{\textsc{q}}].
\end{equation}
This already tells us about the entropy of the spin, since for two dimensional systems, the Von Neumann entropy is just the following monotonically decreasing function of the purity:
\begin{align}\label{EntropyFirst}
  \nn&S(\hat{\rho}_{\textsc{q}}) =\\
  \nn&-\frac{1+\sqrt{2\mathcal{P}(\hat{\rho}_{\textsc{q}})-1}}{2} \log\left(\frac{1+\sqrt{2\mathcal{P}(\hat{\rho}_{\textsc{q}})-1}}{2}\right)\\
  &-\frac{1-\sqrt{2\mathcal{P}(\hat{\rho}_{\textsc{q}})-1}}{2} \log\left(\frac{1-\sqrt{2\mathcal{P}(\hat{\rho}_{\textsc{q}})-1}}{2}\right).
\end{align}
Note that the purity and the entropy are spectral properties of $\hat{\rho}_{\textsc{q}}$, i.e. they only depend on the spectrum of $\hat{\rho}_{\textsc{q}}$ and remain the same independently of any unitary transformation of the state of the system.

 If we write the state $\hat{\rho}^0_{\textsc{q}}$ in terms of its Bloch vector $\bm{u}$ as in equation \eqref{Bloch}, then the initial purity of the spin is just
 \begin{align}
      {\mathcal{P}} (\bm u)=\frac{1+\abs{\bm u}^2}{2},
 \end{align}
where $\abs{\cdot}$ denotes the 3-dimensional Euclidian norm. The purity after the interaction becomes

\begin{align}\label{PurityFirst}
     {\mathcal{P}} \left(\bm{\mathcal{E}}(\bm u)\right)=\frac{1+\abs{\bm u}^2+(1-e^{-4\braket{\hat{O}^2({t}_0)}_{\hat{\rho}_{\textsc{e}}}})\abs{\bm{u}\times\bm{r}(t_0)}^2}{2}
\end{align}

 In the limit $\braket{\hat{O}^2({t}_0)}_{\hat{\rho}_{\textsc{e}}}\to\infty$, the purity reaches a minimum value. Trivially from \eqref{EntropyFirst}, we can write the von-Neumann entropy after the interaction with the environment simply by substituting \eqref{PurityFirst}. In the case where the initial state of the spin-environment system is a product of pure states, this von-Neumann entropy after the interaction gives a measure of entanglement between the spin and the environment generated by the delta-coupling (namely, the entanglement entropy \cite{MR1796805}), which takes the form
\begin{widetext}
\begin{align}
    \nn &S_{\text{Ent}}=S(\mathcal{E}_{t_0}[\ket{\varphi}\!\bra{\varphi}_{\textsc{q}}^{0}])\\
    \nn &= -\frac{1+\sqrt{1+(1-e^{-4\braket{\hat{O}^2({t}_0)}_{\hat{\rho}_{\textsc{e}}}})\abs{\bm{u}\times\bm{r}(t_0)}^2}}{2}
    \log\left(\frac{1+\sqrt{1+(1-e^{-4\braket{\hat{O}^2({t}_0)}_{\hat{\rho}_{\textsc{e}}}})\abs{\bm{u}\times\bm{r}(t_0)}^2}}{2}\right)\\
   &- \frac{1-\sqrt{1+(1-e^{-4\braket{\hat{O}^2({t}_0)}_{\hat{\rho}_{\textsc{e}}}})\abs{\bm{u}\times\bm{r}(t_0)}^2}}{2}
    \log\left(\frac{1-\sqrt{1+(1-e^{-4\braket{\hat{O}^2({t}_0)}_{\hat{\rho}_{\textsc{e}}}})\abs{\bm{u}\times\bm{r}(t_0)}^2}}{2}\right),
\end{align}
\end{widetext}
where $\hat{\rho}_{\textsc{q}}^0=\ket{\varphi}\!\bra{\varphi}_{\textsc{q}}^{0}$ is the initial state of the spin. In summary, the increase of entropy is maximum when the spin state's Bloch vector is perpendicular to $\bm r (t)$ in the Hamiltonian \eqref{hamiltonian} at $t_0$. Note, further, that the cross product of the initial state's Bloch vector with $\bm r (t_0)$ admits an expression in terms of the commutator of the state, that is
 \begin{align}
     \abs{\bm{u}\times\bm{r}(t_0)}=\frac{\norm{[\hat{\rho}_{\textsc{q}},\bm{r}(t_0)\cdot\bm{\hat{\sigma}}]}_1}{4}.
 \end{align}
Here $\norm{\cdot}_{1}$ denotes the trace norm of the argument, i.e. $\norm{\hat A}_{1}=\text{tr}|\hat A|$.

We conclude this section with a brief summary of our results for the single-delta interaction in our spin-boson model: We have fully characterized the partial state of the spin after the delta-interaction and we also have analyzed the specific role played by first two moments of the environment's observable $\hat{O}$ to the spin's dynamics. Moreover, we have characterized the entropy of the spin after the interaction with the field. Furthermore, under the assumption of initially pure states on the environment and the spin, we have analyzed the entanglement generated by interaction between the environment and the spin. Remarkably, we have shown that an interaction that is instantaneous in time can indeed generate entanglement. This entanglement is monotonically increasing with the strength of the fluctuations of the environment at the interaction time $t_0$, encoded in the second moment $\braket{\hat{O}^2({t}_0)}_{\hat{\rho}_{\textsc{e}}}$. Moreover, this entanglement is also increasing with the degree of non-commutation of the spin's initial state with the interaction Hamiltonian, encoded in $\norm{[\bm{r}(t_0)\cdot\bm{\hat{\sigma}} ,\hat{\rho}_{\textsc{q}}]}_1$.

\section{From singular interactions to a collision model}\label{muchas}

In this section we analyze the partial state of the spin after an arbitrary number of delta-like interactions.

Let us first make some preliminary comments. The situation where a bosonic environment interacts with a quantum system repeatedly has been considered to model simple open dynamics \cite{PhysRevA.94.032126}. If one were to think as an environment (say air) interacting with a quantum system (say a qubit), it is not uncommon to hear descriptions of thermalization that involve the following argument: A molecule of air hits the system (interacts for a short time) then leaves and gets lost in the environment. Then a new, fresh, molecule of air hits the system again, and the process repeats indefinitely. If we were to take such a picture seriously, questions would arise about the relevance of the correlations of the environment on the emergent open dynamics: the two different molecules are randomly chosen from air and therefore the two strong-point interactions would not be correlated at all. In other words: an argument could perhaps be made that since the qubit does not interact with the field for long times, the correlations in the environment do not affect the qubit's final state. This would mean that the result of two interactions would be analogous to apply the channel \eqref{CPTP1d} twice at two different times. These are the  situations are commonly considered in collision models, more concretely in set-ups of ancillary bombardment \cite{PhysRevA.97.052120}. In these models, the dynamics is modeled as a sequence of interactions with auxiliary ancillas that are removed after each interaction.  In this section we will analyze the role of time-correlations in the environment and how they affect the spin's response. We will see the exact approximations in which ancillary bombardment will emerge from the general model we describe, and when such approximations are not good and the correlations in the environment do affect the emergent spin dynamics.

We will proceed to some extent in a similar way to section \ref{single}. We can particularize the quantum channel \eqref{CPTP} to a set of $N$ repeated delta-like interactions as
\begin{equation}\label{CPTPcol}
    \mathcal{E}[\hat{\rho}_{\textsc{q}}^0]=\text{tr}_{E}\left[ \hat{U}_{t_0...t_N}\hat{\rho}_{\textsc{q}}^0\otimes\hat{\rho}_{\textsc{e}}\;\hat{U}_{t_0...t_N}^{\dagger}\right],
\end{equation}
where
\begin{align}\label{unitaries}
    \hat{U}_{t_0...t_N}=e^{-\ii\bm{r}({t}_N)\cdot\bm{\hat{\sigma}}\otimes\hat{O}({t}_N)}...e^{-\ii\bm{r}({t}_0)\cdot\bm{\hat{\sigma}}\otimes\hat{O}({t}_0)}
\end{align}
with $0<t_0...<t_N<t_c$.
Note that the dynamics implemented by this channel has similarities with some collision models, e.g.  \cite{Filippov:2017aa}. In \cite{Filippov:2017aa}, the environment was modelled with a chain of correlated qubits instead of bosonic degrees of freedom. However, some of their conclusions are expected to hold also in the spin-boson case studied in this paper. More concretely, we expect the interaction with the train of delta-interactions to implement non-Markovian dynamics. 

Following the same scheme as in section \ref{single}, we expand each factor in \eqref{unitaries} in a partial spectral decomposition in the eigenprojectors of the Pauli operators $\bm{r}({t}_i)\cdot\bm{\hat{\sigma}}$.
\begin{align}\label{unitarieschungas}
    \hat{U}_{t_0...t_N}=\prod_{i=0}^N \left(\hat{P}^+(t_i)\otimes e^{-\ii\hat{O}({t}_i)}+\hat{P}^-(t_i)\otimes e^{\ii\hat{O}({t}_i)}\right),
\end{align}
where, again,
\begin{align}
    \hat{P}^{\pm}(t_i)=\frac{\openone\pm\bm{r}({t_i})\cdot\bm{\hat{\sigma}}}{2}.
\end{align}
Rearranging terms in Eq. \eqref{unitarieschungas} we can write it as a sum:
\begin{align}\label{unitaries2}
    \nn&\hat{U}_{t_N...t_0}\\
    &=\sum_{\{s_i\}=\pm1}\hat{P}^{s_N}(t_N)...\hat{P}^{s_0}(t_0)\otimes e^{-\ii s_N\hat{O}({t}_N)}...e^{-\ii s_0\hat{O}({t}_0)}.
\end{align}
Using \eqref{unitaries2}, it can be shown that the channel \eqref{CPTPcol} acquires the form
\begin{align}\label{CPTP2}
&\mathcal{E}_{t_N...t_0}[\hat{\rho}_{\textsc{q}}^0]\\
&=\!\!\!\!\!\sum_{\!\{s_i,s_i'\}=\pm1}\!\!\!\!\!\gamma_{_{\{s_i,s'_i\}}}\hat{P}^{s_N}({t}_N)\dots\hat{P}^{s_0}({t}_0)\hat{\rho}_{\textsc{q}}^0\hat{P}^{s'_0}({t}_0)\dots\hat{P}^{s'_N}({t}_N)\nn
\end{align}
where
\begin{align}\label{gammas1}
    \gamma_{_{\{s_i,s'_i\}}}=\braket{e^{\ii s'_0\hat{O}({t}_0)}\dots e^{\ii s'_N\hat{O}({t}_N)}e^{-\ii s_N\hat{O}({t}_N)}\dots e^{-\ii s_0\hat{O}({t}_0)}}.
\end{align}
By applying the Weyl relations \eqref{Weyl} recursively, it follows that the coefficients $\gamma_{_{\{s_i,s'_i\}}}$ fulfill
\begin{align}\label{gammas2}
   \nn & \gamma_{_{\{s_i,s'_i\}}}=\braket{e^{\ii\sum_{i=0}^N(s'_i-s_i)\hat{O}(t_i)}}_{\hat{\rho}_{\textsc{e}}}\\
   &\times e^{\frac{1}{2}\sum_{i=1}^{N}\sum_{j=1}^{i-1}(s'_i-s_i)(s'_j+s_j)\mathcal{C}(t_i,t_j)}.
\end{align}
Expression \eqref{gammas2} can be computed for a Gaussian state in terms of the two-point correlator evaluated in pairs of times $t_0...t_N$. Recalling that $\mathcal{C}(t_i,t_j)=\langle[\hat O(t_i),\hat O(t_j)]\rangle$ and considering equation \eqref{mierdasgaussianas2}, we can then write
\begin{widetext}
\begin{align}\label{gaussiangamma}
    \nn\gamma_{_{\{s_i,s'_i\}}}&=\braket{e^{\ii s'_0\hat{O}({t}_0)}...e^{\ii s'_N\hat{O}({t}_N)}e^{-\ii s_N\hat{O}({t}_N)}...e^{-\ii s_0\hat{O}({t}_0)}}_{\hat{\rho}_{\textsc{e}}}=
    e^{\ii\sum_{i=0}^N(s'_i-s_i)\braket{\hat{O}(t_i)}_{\hat{\rho}_{\textsc{e}}}}e^{-\frac{1}{2}\sum_{i=0}^N(s'_i-s_i)^{2}\braket{(\hat{O}(t_i)-\braket{\hat{O}(t_i)}_{\hat{\rho}_{\textsc{e}}}\openone)^{2}}_{\hat{\rho}_{\textsc{e}}}}\\
   & \times e^{-\sum_{i=0}^{N}(s_i-s'_i)\sum_{j=0}^{i-1}s_j\braket{(\hat{O}(t_i)-\braket{\hat{O}(t_i)}_{\hat{\rho}_{\textsc{e}}}\openone)(\hat{O}(t_j)-\braket{\hat{O}(t_j)}_{\hat{\rho}_{\textsc{e}}}\openone)}_{\hat{\rho}_{\textsc{e}}}-s'_j\braket{(\hat{O}(t_j)-\braket{\hat{O}(t_j)}_{\hat{\rho}_{\textsc{e}}}\openone)(\hat{O}(t_i)-\braket{\hat{O}(t_i)}_{\hat{\rho}_{\textsc{e}}}\openone)}_{\hat{\rho}_{\textsc{e}}}},
\end{align}
\end{widetext}
or, if we consider the special case of even Gaussian states
\begin{align}\label{gaussiangammaeven}
    \nn&\gamma_{_{\{s_i,s'_i\}}}=
   e^{-\frac{1}{2}\sum_{i=0}^N(s'_i-s_i)^{2}\braket{\hat{O}^{2}(t_i)}_{\hat{\rho}_{\textsc{e}}}}\\
   &\times e^{-\sum_{i=0}^{N}(s_i-s'_i)\sum_{j=0}^{i-1}s_j\braket{\hat{O}(t_i)\hat{O}(t_j)}_{\hat{\rho}_{\textsc{e}}}-s'_j\braket{(\hat{O}(t_j)\hat{O}(t_i)}_{\hat{\rho}_{\textsc{e}}}}.
\end{align}
Note that if 
\begin{equation}
\braket{(\hat{O}(t_i)-\braket{\hat{O}(t_i)}_{\hat{\rho}_{\textsc{e}}}\openone)(\hat{O}(t_j)-\braket{\hat{O}(t_j)}_{\hat{\rho}_{\textsc{e}}}\openone)}_{\hat{\rho}_{\textsc{e}}}=0
\end{equation}
for $t_i\neq t_j$ (i.e. when the spin is coupled to observables that are not correlated at the different times under consideration) the channel may be written as
\begin{align}\label{CPTPancbom}
    \mathcal{E}_{t_N...t_0}[\hat{\rho}_{\textsc{q}}^0]=\mathcal{E}_{t_N}\circ...\circ\mathcal{E}_{t_0}[\hat{\rho}_{\textsc{q}}^0],
\end{align}
where $\circ$ denotes composition of channels.

This is an important point: we see that the general delta-like interaction  \eqref{CPTPcol} becomes ancillary bombardment (Eq.\eqref{CPTPancbom}) precisely when time-correlations in the environment vanish and thus the bombarding ancillas have no information of what the spin did to previous ancillas. This suggests that short-burst interactions of a system with its environment can indeed model the usual `thermal reservoir' picture where the interaction with the system does not appreciably modify the environment as long as the time correlations of the environment decay sufficiently fast in time. Of course this point has been studied extensively in the literature of open quantum systems \cite{rivas2012open}. 


As an interesting note, we notice that the existence of a self-correlation time scale can help us determine the fixed point of the evolution of the spin in relaxation processes. Namely, consider that a process consisting of $N$ interactions of the system with its environment, and then the environment is left to relax to its initial state. This process (the sequence of interactions plus relaxation) can be repeated $M$ times, then the channel over the spin becomes
\begin{align}
    \mathcal{E}[\hat{\rho}_{\textsc{q}}^0]= \mathcal{E}_{t_N...t_0}^{M}[\hat{\rho}_{\textsc{q}}^0]
\end{align}
where $\mathcal{E}_{t_N...t_0}^{M}$ stands for applying the channel $\mathcal{E}_{t_N...t_0}$ $M$ times.

Since the channel implements an affine transformation over on the Bloch space, it can be written as
\begin{align}
\bm{\mathcal{E}}_{t_N...t_0}(\bm u)=A_{t_N...t_0}\bm{u}+\bm{b}_{t_N...t_0},
\end{align}
after $M$ interactions, the transformation over the Bloch space becomes
\begin{align}
\nn\bm{\mathcal{E}}_{t_N...t_0}^{M}(\bm u)&=A^{M}_{t_N...t_0}\bm{u}\\
&+(\openone-A_{t_N...t_0})^{-1}(\openone-A_{t_N...t_0}^{M-1})\bm{b}_{t_N...t_0}.
\end{align}
Since $\bm{\mathcal{E}}_{t_N...t_0}$ is a completely positive channel, the transformation always remains in the Bloch sphere. Namely, since $\bm{\mathcal{E}}_{t_N...t_0}$ is a contractive map, it always has a fixed point, say $\bm{u}_f$, that may be reached by taking the limit $M\to \infty$. Indeed, this fixed point is
\begin{align}
    \bm{u}_f=(\openone-A_{t_N...t_0})^{-1}\bm{b}_{t_N...t_0}.
\end{align}

\section{Double interaction}\label{double}

In this section we analyze in full detail the properties of the channel induced by a double delta interaction.
The case of two fast interactions is interesting in its own right, since it illustrates the role of the time-correlations between interactions with insightful analytic results.

Let us consider even Gaussian states for simplicity. Particularizing equation \eqref{CPTP2} for two interactions at two times $t_0, t_1$ such that $0<t_0<t_1<t_c$, and for an environment's state that fulfills \eqref{gaussiangamma}, the expression for the density matrix after two interactions is

\begin{align}\label{2deltacutre}
\nn&\mathcal{E}_{t_1,t_0}[\hat{\rho}_{\textsc{q}}^0]=\!\!\!\!\!\!\!\!\!\!\!\!\!\!\!\sum_{s_1,s'_1,s_0,s'_0=1,-1}\!\!\!\!\!\!\!\!\!\!\!\!\!\!\gamma_{s_1,s'_1,s_0,s'_0}\hat{P}^{s_1}({t}_1)\hat{P}^{s_0}({t}_0)\hat{\rho}_{\textsc{q}}^0\hat{P}^{s'_0}({t}_0)\hat{P}^{s'_1}({t}_1)\\
\nn&=\!\!\!\!\!\!\!\!\!\!\!\! \sum_{s_1,s'_1,s_0,s'_0=1,-1}\!\!\!\!\!\!\!\!\!\!\!\! e^{-\frac{(s_1-s'_1)^2}{2}\braket{\hat{O}^2(t_1)}_{\hat{\rho}_{\textsc{e}}}}\;e^{-\frac{(s_0-s'_0)^2}{2}\braket{\hat{O}^2(t_0)}_{\hat{\rho}_{\textsc{e}}}}\\
\nn&\times e^{-(s_1-s'_1)\left(s_0\braket{\hat{O}(t_1)\hat{O}(t_0)}_{\hat{\rho}_{\textsc{e}}}-s'_0\braket{\hat{O}(t_0)\hat{O}(t_1)}_{\hat{\rho}_{\textsc{e}}}\right)}\\
&\times \hat{P}^{s_1}({t}_1)\hat{P}^{s_0}({t}_0)\hat{\rho}_{\textsc{q}}^0\hat{P}^{s'_0}({t}_0)\hat{P}^{s'_1}({t}_1).
\end{align}

After performing the sums and substituting \eqref{gaussiangammaeven}, expression \eqref{2deltacutre} takes the form
\begin{widetext}
\begin{align}\label{CPTP2d}
\nn\mathcal{E}_{t_1,t_0}[\hat{\rho}_{\textsc{q}}^0]&=\frac{1}{4}\left[(\text{Id} + \mathcal{U}(t_1))\circ(\text{Id} + \mathcal{U}(t_0))[\hat{\rho}_{\textsc{q}}^0]\right)+
\frac{e^{-2\braket{\hat{O}^2(t_0)}_{\hat{\rho}_{\textsc{e}}}}}{4}\left((\text{Id} + \mathcal{U}(t_1))\circ(\text{Id} - \mathcal{U}(t_0))[\hat{\rho}_{\textsc{q}}^0]\right)\\
\nn&+\frac{e^{-2\braket{\hat{O}^2(t_1)}_{\hat{\rho}_{\textsc{e}}}}\cos\left(4\Im  \braket{\hat{O}(t_1)\hat{O}(t_0)}_{\hat{\rho}_{\textsc{e}}}\right)}{4}\left((\text{Id} - \mathcal{U}(t_1))\circ(\text{Id} + \mathcal{U}(t_0))[\hat{\rho}_{\textsc{q}}^0]\right)\\
\nn&+\frac{e^{-2\braket{\hat{O}^2(t_1)}_{\hat{\rho}_{\textsc{e}}}}e^{-2\braket{\hat{O}^2(t_0)}_{\hat{\rho}_{\textsc{e}}}}\cosh\left(4\Re  \braket{\hat{O}(t_1)\hat{O}(t_0)}_{\hat{\rho}_{\textsc{e}}}\right)}{4}\left((\text{Id} - \mathcal{U}(t_1))\circ(\text{Id} - \mathcal{U}(t_0))[\hat{\rho}_{\textsc{q}}^0]\right)\\
\nn &+\frac{-\ii e^{-2\braket{\hat{O}^2(t_1)}_{\hat{\rho}_{\textsc{e}}}}\sin\left(4\Im  \braket{\hat{O}(t_1)\hat{O}(t_0)}_{\hat{\rho}_{\textsc{e}}}\right)}{4}\left[\bm{r}({t_1})\!\cdot\!\bm{\hat{\sigma}},\left\{\bm{r}({t_0})\!\cdot\!\bm{\hat{\sigma}},\hat{\rho}_{\textsc{q}}^0\right\}\right]\\
 &-\frac{e^{-2\braket{\hat{O}^2(t_1)}_{\hat{\rho}_{\textsc{e}}}}e^{-2\braket{\hat{O}^2(t_0)}_{\hat{\rho}_{\textsc{e}}}}\sinh\left(4\Re  \braket{\hat{O}(t_1)\hat{O}(t_0)}_{\hat{\rho}_{\textsc{e}}}\right)}{4}\left[\bm{r}({t_1})\!\cdot\!\bm{\hat{\sigma}},\left[\bm{r}({t_0})\!\cdot\!\bm{\hat{\sigma}},\hat{\rho}_{\textsc{q}}^0\right]\right],
\end{align}
\end{widetext}
where $\circ$, again, denotes the composition of channels. We have used that
\begin{align}
    \braket{\hat{O}(t_1)\hat{O}(t_0)}_{\hat{\rho}_{\textsc{e}}}=\braket{\hat{O}(t_0)\hat{O}(t_1)}^{*}_{\hat{\rho}_{\textsc{e}}},
\end{align}
and the properties of exponentials, hyperbolic and trigonometric functions.

 The channel \eqref{2deltacutre}, similarly to the single interaction channel \eqref{CPTP1d}, can be represented as an affine transformation in the Bloch sphere. Namely, 
\begin{align}\label{CPTP2dBloch}
\nn&\bm{\mathcal{E}}_{t_1,t_0}(\bm{u})=\bm{u}_{1,0}\\
\nn &+ e^{-2\braket{\hat{O}^2(t_0)}_{\hat{\rho}_{\textsc{e}}}}(\bm{u}_{1}- \bm{u}_{1,0})\\
\nn&+ e^{-2\braket{\hat{O}^2(t_1)}_{\hat{\rho}_{\textsc{e}}}}\left[\cos\left(4\Im  \braket{\hat{O}(t_1)\hat{O}(t_0)}_{\hat{\rho}_{\textsc{e}}}\right)
(\bm{u}_{0}- \bm{u}_{1,0})\right.\\
\nn&\left.+\sin\left(4\Im  \braket{\hat{O}(t_1)\hat{O}(t_0)}_{\hat{\rho}_{\textsc{e}}}\right)\bm r (t_1)\times\bm r (t_0)\right]\\
\nn& +e^{-2\braket{\hat{O}^2(t_1)}_{\hat{\rho}_{\textsc{e}}}}e^{-2\braket{\hat{O}^2(t_0)}_{\hat{\rho}_{\textsc{e}}}}\\
\nn&\times\left[\cosh\left(4\Re  \braket{\hat{O}(t_1)\hat{O}(t_0)}_{\hat{\rho}_{\textsc{e}}}\right)(\bm{u}-\bm{u}_{1}-\bm{u}_{0}+\bm{u}_{1,0})\right.\\
&\left.+\sinh\left(4\Re  \braket{\hat{O}(t_1)\hat{O}(t_0)}_{\hat{\rho}_{\textsc{e}}}\right)\bm r (t_1)\times(\bm r (t_0)\times \bm{u})\right],
\end{align}
where we have defined the projections
\begin{align}
    \bm{u}_{0}=(\bm u \cdot\bm r (t_0))\bm r (t_0),
\end{align}
\begin{align}
    \bm{u}_{1}=(\bm u \cdot\bm r (t_1))\bm r (t_1),
\end{align}
and
\begin{align}
    \bm{u}_{1,0}=(\bm u \cdot\bm r (t_0))(\bm r (t_0)\cdot\bm r (t_0))\bm r (t_1).
\end{align}
Note that, since equation \eqref{CPTP2dBloch} defines an affine transformation on the Bloch sphere, it can be written as
\begin{align}
    \bm {\mathcal{E}}_{t_1,t_0}(\bm u)=A\bm u+\bm b,
\end{align}
where $A$ is a matrix and $\bm b$ is a constant vector.
 In order to give an explicit expression for $A$ and $\bm{b}$, we need to choose an orthonormal basis in $\mathbb{R}^3$.  In our current case, a well-suited basis in $\mathbb{R}^3$  is given by the following basis elements
\begin{align}\label{vecbasis}
   \nonumber &\bm{e}_1=\bm r (t_1),& \bm{e}_2=\frac{\bm r (t_0)-(\bm r (t_1) \cdot\bm r (t_0))\bm r (t_1)}{\abs{\bm r (t_1)\times\bm r (t_0)}},\\
    &\bm{e}_3=\frac{\bm r (t_1)\times\bm r (t_0)}{\abs{\bm r (t_1)\times\bm r (t_0)}}.&
\end{align}
Indeed, in this basis and after some formal manipulations, the transformation can be written in a compact form in terms of a few parameters. In this basis the channel takes the form
\begin{align}
    \bm {\mathcal{E}}_{t_1,t_0}(\bm u)=BC\bm u+\bm b,
\end{align}
with 
\begin{align}
B=
    \begin{bmatrix}
     1 &  0& 0 \\
    2\Re(h k^*)      & |h|^2-|k|^2 & 0 \\
 0 & 0 &|h|^2 +|k|^2
\end{bmatrix},
\end{align}
\begin{align}\label{Chunga}
C=
    \begin{bmatrix}
     \alpha^2+g(1-\alpha^2)& \alpha \sqrt{1-\alpha^2}(1-g) & 0 \\
    \alpha \sqrt{1-\alpha^2}(1-g)   & g \alpha^2+(1-\alpha^2) & 0 \\
 0 & 0 &g
\end{bmatrix}
\end{align}
and
\begin{align}
\bm{b}=
    \begin{bmatrix}
     0 \\
     0 \\
  2\Im(h k^*)
\end{bmatrix}.
\end{align}

We have parametrized the transformation in terms of the following quantities
\begin{align}
    \alpha=\bm r (t_1) \cdot\bm r (t_0),
\end{align}
\begin{align}
    g= e^{-2\braket{\hat{O}^2(t_0)}_{\hat{\rho}_{\textsc{e}}}},
\end{align}
\begin{align}\label{hache}
\nn h&=   e^{-\braket{\hat{O}^2(t_1)}_{\hat{\rho}_{\textsc{e}}}} \left(\cosh{(2\braket{\hat{O}(t_1)\hat{O}(t_0)}_{\hat{\rho}_{\textsc{e}}})}\right.\\
&\left.-(\bm{r}({t_1})\cdot\bm{r}({t_0}))\sinh{(2\braket{\hat{O}(t_1)\hat{O}(t_0)}_{\hat{\rho}_{\textsc{e}}}})\right)
\end{align}
and 
\begin{align}\label{ka}
 k&= \abs{\bm r(t_1)\times\bm r(t_0)} e^{-\braket{\hat{O}^2(t_1)}_{\hat{\rho}_{\textsc{e}}}} \sinh{(2\braket{\hat{O}(t_1)\hat{O}(t_0)}_{\hat{\rho}_{\textsc{e}}})}.
\end{align}
As a technical point, notice that Eq. \eqref{CPTP2dBloch} simplifies greatly in the case where the initial Bloch vector of the spin is proportional to $\bm{e}_3$. 
Indeed, in that case
\begin{align}
\nn&\bm{\mathcal{E}}_{t_1,t_0}(\bm{u})=g(|h|^2 +|k|^2)\bm{u}+  2\Im(h k^*)\bm{e}_3\\
\nn&=e^{-2\braket{\hat{O}^2(t_1)}_{\hat{\rho}_{\textsc{e}}}}e^{-2\braket{\hat{O}^2(t_0)}_{\hat{\rho}_{\textsc{e}}}}\\
\nn&\times\left[\cosh\left(4\Re  \braket{\hat{O}(t_1)\hat{O}(t_0)}_{\hat{\rho}_{\textsc{e}}}\right)\right.\\ 
\nn &-\left.(\bm r (t_1)\cdot\bm r (t_0))\sinh\left(4\Re  \braket{\hat{O}(t_1)\hat{O}(t_0)}_{\hat{\rho}_{\textsc{e}}}\right) \right]\bm u\\
&+ e^{-2\braket{\hat{O}^2(t_1)}_{\hat{\rho}_{\textsc{e}}}}
\sin\left(4\Im  \braket{\hat{O}(t_1)\hat{O}(t_0)}_{\hat{\rho}_{\textsc{e}}}\right)\bm r (t_1)\times\bm r (t_0).
\end{align}
This situation arises, e.g., when $\bm{h}$ and $\bm{u}$ are parallel, and both perpendicular to $\bm r(t)$ for all $t$. This is an interesting case since the initial state of the spin may be prepared in the eigenbasis of the free Hamiltonian (think for example a free Hamiltonian proportional to $\hat \sigma_z$ and an interaction Hamiltonian in the subspace spanned by $\hat \sigma_x,\hat \sigma_y$) as it is common in experimental settings.

The general two-interaction channel is non-unital: it is straightforward to check from equation \eqref{CPTP2d} that its action on the identity is
\begin{align}
   \nn \mathcal{E}_{t_1,t_0}[\openone]=&\openone+e^{-2\braket{\hat{O}^2(t_1)}_{\hat{\rho}_{\textsc{e}}}}\\
   &\nn\times\sin\left(4\Im  \braket{\hat{O}(t_1)\hat{O}(t_0)}_{\hat{\rho}_{\textsc{e}}}\right)\;\bm{r}({t_1})\times\bm{r}({t_0})\cdot \bm{\sigma}.
\end{align}
Therefore, the channel can increment the purity of the system. Notice that a quantum channel on a finite-dimensonal quantum system is unital if and only if it can decrease the system's entropy as discussed in \cite{Lindblad_1973} and shown here explcitly in Appendix \ref{entropygain}. 

The purity of the state after the interaction is is given by the quadratic function
\begin{align}
  \nn  {\mathcal{P}} \left(\bm{\mathcal{E}}_{t_1,t_0}(\bm u)\right)=\frac{1+\norm{BC\bm{u}}^2+2\bm{b}\cdot BC\bm{u}+\norm{\bm{b}}^2}{2}=\\
    \frac{1+\norm{BC\bm{u}}^2}{2}+g(|h|^2 +|k|^2)\bm{e}_3\cdot\bm{u}+2\Im(h k^*)^2
\end{align}



\section{Many interactions and emergence of pure dephasing}\label{dephasing}

\subsection{Coupling to commuting observables}

In this section we analyze the particular case in which the Pauli observable coupling to the bosonic environment commutes with itself at the different times under consideration, that is,
\begin{equation}
    [\bm{r}(t_i)\cdot \bm \hat\sigma,\bm{r}(t_j)\cdot \bm \hat\sigma]=\Big(\bm{r}(t_i)\times\bm{r}(t_j)\Big)\cdot \hat{\bm\sigma}=0.
\end{equation}

We will show that the dynamics of the spin is, to a large extent, captured by the single interaction channel described in section \ref{single}. First, note that the Pauli observable in the interaction Hamiltonian \eqref{hamiltonian}, characterized by $\bm{r}(t)$ in this case is such that $\bm{r}(t_i)\times\bm{r}(t_j)=0$. Because of this, the the procedure described in previous sections is considerably less involved. The reason is that the channel \eqref{CPTPcol} is just a phase damping in a time-independent basis (in the interaction picture). This follows directly from the unitary operator \eqref{unitaries2}.  Since $[\bm{r}(t_i)\cdot \bm \hat{\sigma},\bm{r}(t_j)\cdot \bm \hat{\sigma}]=0$, all the relevant Pauli operators admit a common spectral decomposition of the form 
\begin{align}\label{commonsd}
    \bm{r}(t_i)\cdot \bm{\hat \sigma}=f(t_{i})(\hat{P}^{+}-\hat{P}^{-}),
\end{align}
where $f(t_{i})=\pm 1$. Let us notate the projectors $\hat P^{s}$, with $s=\pm1$ and
\begin{align}\label{errequeerre}
    \hat P^{s}=\frac{\openone+s\,\bm r\cdot\hat{\bm\sigma}}{2},
\end{align}
where $\bm{r}$ is a unit vector independent on time.
 Since they are orthogonal projectors, they fulfill
\begin{align}
    \hat{P}^{s}\hat{P}^{s'}=\delta_{s s'}\hat{P}^{s},
\end{align}
and equation \eqref{unitaries2} simplifies to 
\begin{align}\label{unitariesmany}
   \nn\hat{ U}_{t_N..t_0}&=\hat{P}^+\otimes e^{-\ii f({t}_N)\hat{O}({t}_N)}\dots e^{-\ii f({t}_0)\hat{O}({t}_0)}\\
   &+\hat{P}^-\otimes e^{\ii f({t}_N)\hat{O}({t}_N)}\dots e^{\ii f({t}_0)\hat{O}({t}_0)}.
\end{align}
Let us focus our attention on the role of the time-ordering in equation \eqref{unitariesmany}.  Again, by recursively applying the Weyl relations \eqref{Weyl}, the unitary evolution of the whole system can be written as 
\begin{align}
   &\nn\hat{ U}_{t_N..t_0}\\
   \nn&=\hat{P}^+\otimes  e^{-\frac{1}{2}\sum_{i=1}^{N}\sum_{j=1}^{i-1}f({t}_i)f({t}_j)\mathcal{C}(t_i,t_j)} e^{-\ii\sum^{N}_{i=0}f({t}_i)\hat{O}({t}_i)}\\
   \nn&+\hat{P}^-\otimes e^{-\frac{1}{2}\sum_{i=1}^{N}\sum_{j=1}^{i-1}f({t}_i)f({t}_j)\mathcal{C}(t_i,t_j)} e^{\ii\sum^{N}_{i=0}f({t}_i)\hat{O}({t}_i)}\\
   \nn&=e^{-\frac{1}{2}\sum_{i=1}^{N}\sum_{j=1}^{i-1}f({t}_i)f({t}_j)\mathcal{C}(t_i,t_j)}\\
   &\times \left(\hat{P}^+\otimes   e^{-\ii\sum^{N}_{i=0}f({t}_i)\hat{O}({t}_i)}+\hat{P}^-\otimes e^{\ii\sum^{N}_{i=0}f({t}_i)\hat{O}({t}_i)}\right).
\end{align}
Since $\mathcal{C}(t_i,t_j)$ is pure imaginary, the time ordering of the spin interactions through Pauli operators only implements a global phase, which is physically irrelevant. In contrast, for higher-dimensional systems (and also for more general spins couplings) the time-ordering of the delta-interactions will be relevant in the system's dynamics. Namely, the time sequence of the couplings will matter when the observable that couples to the bosonic environment is not unitary. Overall, different time orderings will induce different phases associated with different eigenvalues of the system's observable coupling at each time. Hence, different time orderings yield different relative phases in the channel. For the explicit proof of this statement and a detailed analysis of the higher-dimensional case for arbitrary switchings (not only deltas), see Appendix \ref{puredephasing}.

Now we proceed to the analysis of our concrete model. In this situation the channel over the spin can be written as
\begin{align}
   \nn\mathcal{E}[\hat{\rho}_{\textsc{q}}^0]&=\hat{P}^-\hat{\rho}_{\textsc{q}}^0\hat{P}^- +  \hat{P}^+\hat{\rho}_{\textsc{q}}^0\hat{P}^+\\
   & + \gamma_{+-}\hat{P}^+\hat{\rho}_{\textsc{q}}^0\hat{P}^-+ \gamma_{+-}^* \hat{P}^-\hat{\rho}_{\textsc{q}}^0\hat{P}^+,
\end{align}
or
\begin{align}
  \nn   \mathcal{E}_{t_n,\dots t_0}[\hat{\rho}_{\textsc{q}}^0]&=\frac{1}{2}(\text{Id}+\mathcal{U})[\hat{\rho}_{\textsc{q}}^0]\\
  & + \frac{\Re \gamma_{+-}}{2}(\text{Id}-\mathcal{U})[\hat{\rho}_{\textsc{q}}^0]
  +\frac{\ii\Im \gamma_{+-}}{2}\left[\bm{r}\!\cdot\!\bm{\hat{\sigma}},\hat{\rho}\right],
\end{align}
where 
\begin{align}\label{gammadeph}
    \gamma_{+-}=\braket{e^{-2\ii\sum^{N}_{i=0}f({t}_i)\hat{O}({t}_i)}}.
\end{align}

From equation \eqref{gaussiangamma}, setting with $s_0...s_N=1$ and $s'_0...s'_N=-1$ we see that for an even Gaussian environment's state, it follows that the phase damping channel coefficient $\gamma _{+-}$ takes the form
\begin{align}
\nn&\gamma _{+-}=e^{-2\braket{\left(\sum_{i=0}^N f({t}_i)\hat{O}(t_i)\right)^2}_{\hat{\rho}_{\textsc{e}}}}\\
&= e^{-2\sum_{j=0}^{N}\sum_{i=0}^N f({t}_i)f({t}_j)\braket{\hat{O}(t_i)\hat{O}(t_j)}_{\hat{\rho}_{\textsc{e}}}}.
\end{align}
 From this, we see that the time-correlations between the environment observables the system couples to at each interaction dictate the magnitude of the damping. 
 
 Let us draw our attention to an interesting feature of this multiple-delta coupling. If a system is `kicked' by an environment several times, one would expect that the action of the system on the environment will start backreacting on the system after several interactions. Consequently, one may expect the system's channel to encode information about the changes on the environment. However we see that this is not the case: since the time sequence of the interactions has no physical effect it is not possible for them to encode anything about processes on the environment. Notice the key role of the irrelevance of time-ordering for this feature. Because of this, the lack of information in the spin dynamics about the effect of the coupling on the environment is a peculiarity of the unitarity of the observable of the system that couples to the bosonic environment.

\subsection{Cases of interest: When do coupling observables commute?}

The situation described above (that the system couples repeatedly to commuting observables of the environment) can be achieved in several physical scenarios. First, one could consider that the Bloch vector describing the interaction Hamiltonian is parallel to the one describing the free Hamiltonian, i.e. $\bm \alpha=\pm\bm h$. In that case, it is clear that $\bm r=\bm\alpha$ is time independent, thus commuting with itself at all times. A similar situation may be achieved by setting $\Omega=0$, in such a way that the spin is not subjected to free evolution. 

An alternative way to reach time independence in the coupling observables in $\bm r$ comes from noticing that $\bm r$ is a periodic function of time.  As it was already noted in \cite{PhysRevD.98.105011}, applying periodic interactions with a frequency multiple to the natural frequency of the spin $\Omega$, leads to an effectively time-independent Bloch vector $\bm r$.  Indeed, by inspecting equation \eqref{erredete}, we realize that
\begin{align}
    \bm r(2\pi n\Omega^{-1}+t_0)=\bm r(t_0),
\end{align}
 with $n\in\mathbb{N}$. Hence, the function $f(t_n)=1$ for all $n$.

 Further, when the Bloch vector $\bm\alpha$ is perpendicular to the free Hamiltonian, i.e. $\bm\alpha\cdot\bm h=0$, it holds that
 \begin{align}
    \bm r(\pi n\Omega^{-1}+t_0)=(-1)^n\bm r(t_0),
\end{align}
therefore, $f(t_n)=(-1)^n$.

\section{Divisibility and memory effects}\label{sec:Marcopolo}

In this section we study non-Markovian effects in the dynamics of the partial state of the spin after a series of several interactions. We first focus on the case of two delta-interactions but in full generality. After that, we will discuss the non-Markovian character of the interaction for an arbitrary number of delta-kicks in pure dephasing scenarios (i.e., synchronized interactions or degenerate spins).

Given the CPTP channel \eqref{CPTP2d}, one could wonder whether it exhibits memory effects. Fundamental notions in this regard, such as divisibility---or more concrete and rigorous notions of memory in open dynamics---have been extensively studied in the literature (see e.g., \cite{Rivas_2014}). However, we provide a brief introduction here. 

Let the state of a quantum system at a certain time be $\hat\rho_{t_0}$. Then, the state of the system at a later time $t_1$ can be written as
\begin{align}
    \hat\rho_{t_1}=\mathcal{E}_{t_1,t_0}{[\hat\rho_{t_0}]},
\end{align}
where $\mathcal{E}_{t_1,t_0}$ is a CPTP channel.
At a later time $t_2$, the system is in a state
\begin{align}
    \hat\rho_{t_2}=\mathcal{E}_{t_2,t_0}{[\hat\rho_{t_0}]}.
\end{align}
Note that, at least formally, we can always write 
\begin{align}
    \hat\rho_{t_2}=\Theta_{t_2,t_1}{[\hat\rho_{t_1}]},
\end{align}
where $\Theta_{t_2,t_1}$ is known as the transition map. $\Theta_{t_2,t_1}$ connects the state of a system at different times. Let us introduce the notion of divisibility.
\begin{defi}
A dynamical map $\mathcal{E}$ is P-divisible if its transition maps $\Theta$ are positive linear maps.
\end{defi}
Therefore, for a P-divisible process we have that the transition maps are also dynamical maps, and it makes sense to talk about `intermediate states' in a given process. When a process is divisible, we can write
\begin{align}
    \hat\rho_{t_2}=\mathcal{E}_{t_2,t_0}{[\hat\rho_{t_0}]}=\mathcal{E}_{t_2,t_1}[\mathcal{E}_{t_1,t_0}{[\hat\rho_{t_0}]]}.
\end{align}
Note that some authors do not demand the trace preserving condition since they attempt to include operations such as measurements in the definition of dynamical map~\cite{}. We will not enter in such considerations, so here P-divisibility implies also trace preservation. 

Note that the notion of intermediate state, does not require that the transition maps are completely positive, but just positive. This is why we have defined P-divisibility (the $P$ standing for `positive') instead of CP-divisibility (standing for `completely positive'). However, notice that while the transition maps do not need to be completely positive, we have good reasons to still demand that the full dynamical map $\mathcal{E}$ is compeltely positive. Indeed, a map that is positive does not remain necessarily positive when it acts on a higher dimensional Hilbert space, even when they act trivially in the extended Hilbert space. It can be shown that the presence of entanglement in the higher dimensional space can break the positivity of dynamical maps even if the transition maps are positive \cite{paulsen_2003,MR1796805}.  

We can now present a common way of introducing the notion of Markovianity in quantum systems, which is to define Markovianity as CP-divisibility, established from the the following definition \cite{Rivas_2014}:
\begin{defi}
 A dynamical map is CP-divisible if its transition maps are themselves completely positive. A dynamical map is Markovian if it is CP-divisible.
\end{defi}

Whereas in the classical set-up the relation of the Markovian property with the memory of the dynamics is explicit, the link between divisibility and  memory effects in the dynamics is not so straightforward. In order to link the two notions, we have to analyze a characteristic of divisible maps, also present in divisible, classical stochastic processes, called \textit{contractive property}. 

First, consider the following one-shot discrimination problem. We are given two different states $\hat\rho_1$ and $\hat\rho_2$, one with probability $p$ and the the other with probability $1-p$. Then we perform a measurement in the system with the aim to discriminate if the state is either $\hat\rho_1$ or $\hat\rho_2$. In order to do so, we define a thought experiment modelled by a positive operator valued measure (POVM) characterized by two POVM elements $\hat T$ and $\openone-\hat T$. If the result of the experiment is the value associated with the element $\hat T$, we conclude that the state is $\hat\rho_1$, and $\hat\rho_2$ otherwise. Then we misidentify the state with average probability
\begin{align}
 P_{\text{fail}}=   p\,\text{tr}(\hat\rho_1 (\openone-\hat T)) +(1-p)\text{tr}(\hat\rho_2 \hat T).
\end{align}
Then,there is an experiment setup (a choice of $\hat T$) such that this probability is minimized. It can be shown \cite{Rivas_2014} that for such $\hat T$ the minimum probability of misidentifying the state is given by
\begin{align}
     P_{\text{fail}}=\frac{1-\norm{p\hat\rho_1-(1-p)\hat\rho_2}_{1}}{2}.
\end{align}
 We can interpret the probability of failing in a one-shot discrimination problem as a measure of information that we have about the system. We will now relate this measure of information about the system with the divisibility of dynamical maps. It can be shown that if a map is positive and trace preserving, then it is contractive, that is
\begin{align}
    \norm{\mathcal{E}[\hat A]}_{1}\leq\norm{\hat A}_{1},
\end{align}
for any operator $\hat A$.
Then, if we have a divisible map
\begin{align}
   \norm{\hat\rho_{t}}_{1} =\norm{\mathcal{E}_{t,t_0}[\hat\rho_{t_0}]}_{1}=\norm{\mathcal{E}_{t,t'}[\hat\rho_{t'}]}_{1}\leq\norm{\hat\rho_{t'}}_{1}
\end{align}
if $t>t'$. Therefore, in any Markovian evolution of a quantum system the information, understood as the probability of success in a one-shot identification problem, decreases monotonically with time. The link with the concept of memory is that under Markovian dynamics there cannot be any revival of the information (understood as above) about the system in previous times. It is, hence, common in the literature to associate these revivals with non-Markovian effects. When these dynamics are associated with a system-enviroment interaction such revivals are usually interpreted as a {\it back-flow of information} from the environment to the system \cite{Chruscinski:2011aa}. 

Let us now refocus our attention on the spin-boson interaction.
As explained above, the non-Markovian character of the spin-boson dynamics relies on determining whether the map $\Theta_{t_1,t_0}$ is a completely positive and trace preserving map, since in this case no information can be gained respect to the state after just one interaction. 

Lack of divisibility in the channel generated by two delta-couplings in the  spin-boson model can be interpreted as the gain of information of applying two interactions. For instance, if the spin starts in a known pure state, after the first interaction it can become correlated with the environment, then the initial information about the spin's state is always partially lost or, in the best case scenario, remain the same. Indeed, the same happens after any number of delta-interactions. However, one may wonder whether we lose more information about the spin's initial state in the case of a single delta interaction or in the case of multiple delta-couplings to the boson environment. The answer to this question is intrinsically related with the divisibility of the channel. To see how, let us illustrate this with a concrete analysis in the case of two delta interactions. 

Consider the channel \eqref{CPTP2d}. First we determine if the transition map, given implicitly by the relation 
\begin{align}
    \mathcal{E}_{t_1,t_0}[\hat{\hat\rho}_{\textsc{q}}^0]=\Theta_{t_1,t_0}\circ\mathcal{E}_{t_0}[\hat{\hat\rho}_{\textsc{q}}^0],
\end{align}
can be written in closed form. In orther to do so, consider the following map
\begin{align}
  \nn   \mathcal{E}^{-1}_{t_0}[\hat{\rho}_{\textsc{q}}^0]&=\frac{1}{2}(\text{Id}+\mathcal{U}(t_0))[\hat{\rho}_{\textsc{q}}^0]\\
  &  +\frac{e^{2\braket{\hat{O}^2({t}_0)}_{\hat{\rho}_{\textsc{e}}}}}{2}(\text{Id}-\mathcal{U}(t_0))[\hat{\rho}_{\textsc{q}}^0].
\end{align}
It is straightforward to see that $\mathcal{E}^{-1}_{t_0}\circ\mathcal{E}_{t_0}=\mathcal{E}_{t_0}\circ\mathcal{E}^{-1}_{t_0}=$Id, since $\mathcal{U}^2(t_0)=\mathcal{U}(t_0)\circ\mathcal{U}(t_0)$=Id. Therefore, the channel $\mathcal{E}_{t_0}$ always has an inverse, in terms of which $\Theta_{t_1,t_0}$ has the expression 
\begin{align}\label{eq:invtrans}
    \Theta_{t_1,t_0}=\mathcal{E}_{t_1,t_0}\circ\mathcal{E}^{-1}_{t_0}.
\end{align}
 In fact, from \eqref{eq:invtrans}, using \eqref{CPTP2d},  the expression of  $\Theta_{t_1,t_0}$  can be written explicitly,
\begin{widetext}
\begin{align}\label{intermap}
\nn&\Theta_{t_1,t_0}[\hat{\rho}_{\textsc{q}}^0]=\mathcal{E}_{t_1}[\hat{\rho}_{\textsc{q}}^0]+e^{-2\braket{\hat{O}^2(t_1)}_{\hat{\rho}_{\textsc{e}}}}\left[-\frac{\sin^2\left(2\Im  \braket{\hat{O}(t_1)\hat{O}(t_0)}_{\hat{\rho}_{\textsc{e}}}\right)}{2}\left((\text{Id} - \mathcal{U}(t_1))\circ(\text{Id} + \mathcal{U}(t_0))[\hat{\rho}_{\textsc{q}}^0]\right)\right.\\
\nn&+\frac{\sinh^2\left(2\Re  \braket{\hat{O}(t_1)\hat{O}(t_0)}_{\hat{\rho}_{\textsc{e}}}\right)}{2}\left((\text{Id} - \mathcal{U}(t_1))\circ(\text{Id} - \mathcal{U}(t_0))[\hat{\rho}_{\textsc{q}}^0]\right)\\
 &\left.\frac{-\ii\sin\left(4\Im  \braket{\hat{O}(t_1)\hat{O}(t_0)}_{\hat{\rho}_{\textsc{e}}}\right)}{4}\left[\bm{r}({t_1})\!\cdot\!\bm{\hat{\sigma}},\left\{\bm{r}({t_0})\!\cdot\!\bm{\hat{\sigma}},\hat{\rho}_{\textsc{q}}^0\right\}\right]-
\frac{\sinh\left(4\Re  \braket{\hat{O}(t_1)\hat{O}(t_0)}_{\hat{\rho}_{\textsc{e}}}\right)}{4}\left[\bm{r}({t_1})\!\cdot\!\bm{\hat{\sigma}},\left[\bm{r}({t_0})\!\cdot\!\bm{\hat{\sigma}},\hat{\rho}_{\textsc{q}}^0\right]\right]\right].
\end{align}
\end{widetext}
Note that the transition map \eqref{intermap} is then just the channel \eqref{CPTP2d} in the limit $\braket{\hat{O}^2(t_0)}_{\hat{\rho}_{\textsc{e}}}\to 0$.
Equation \eqref{intermap} shows that in the particular case where the correlations in the environment vanish, i.e. $\braket{\hat{O}(t_1)\hat{O}(t_0)}_{\hat{\rho}_{\textsc{e}}}=0$, the map $\Theta_{t_1,t_0}$ is completely positive, thereby the channel $\mathcal{E}_{t_1,t_0}$ is CP-divisible.

In more general cases where correlations do not vanish, to study the complete positive character of $\Theta_{t_1,t_0}$ it is more convenient to work in the $\chi-$matrix representation (see appendix \ref{chiap}). 
The map $\Theta_{t_1,t_0}$  can be written as
\begin{align}\label{chichi}
   \Theta_{t_1,t_0}[\hat{\rho}_{\textsc{q}}^0]= \sum_{a b}\bm\chi_{ab} \hat B_a\hat{\rho}_{\textsc{q}}^0\hat B_b^\dagger
\end{align}
where $\bm\chi_{ab}$ is the  $\chi-$matrix. It can be shown \cite{nambu2005matrix} that a map is completely positive if and only if the matrix $\bm\chi_{ab}$ is positive semi-definite. The sufficiency of this condition is clear, if $\bm\chi_{ab}$ is positive semidefinite then it can be diagonalized and all its eigenvalues are positive, then expression \eqref{chichi} can be written in its diagonal form by redefining the basis elements $\hat B_a$, and then it is a Kraus form of the channel.

In order to find the $\chi$-matrix representation for \eqref{intermap}, it is convenient to choose the following  basis of operators:
\begin{align}\label{basis}
   \nn& \hat B_0=\frac{\openone}{\sqrt{2}},\qquad  \hat B_2=\frac{\left(\bm{r}({t_0})-(\bm{r}({t_1})\cdot\bm{r}({t_0}))\bm{r}({t_1})\right)\cdot\bm{\hat{\sigma}}}{\sqrt{2}\abs{\bm r(t_1)\times\bm r(t_0)}}\\
  & \hat B_1=\frac{\bm{r}({t_1})\cdot\bm{\hat{\sigma}}}{\sqrt{2}},\quad \hat B_3= \frac{\bm r(t_1)\times\bm r(t_0)\cdot\bm{\hat{\sigma}}}{\sqrt{2}\abs{\bm r(t_1)\times\bm r(t_0)}}.&
\end{align}
Since $\abs{\bm r(t)}=1$, it can be checked that this set forms an orthogonal basis with respect to the inner product \begin{align}\label{innerp}
    \langle\hat A,\hat B\rangle=\tr(\hat B^\dagger \hat A).
\end{align}


In the basis \eqref{innerp}, and after some formal manipulations, the $\chi-$matrix can be written explicitly as
\begin{align}
\bm{\chi}=
    \begin{bmatrix}
     1+|h|^2  &  0& 0 &  \ii h^* k  \\
    0       & 1-|h|^2 & h^* k & 0 \\
   0 &h k^* &  - |k|^2  &0\\
  -\ii h k^*   & 0 & 0 &   |k|^2 
\end{bmatrix}
\end{align}\
where we recall that the parameters $h,k$ are  given by equations \eqref{hache}, \eqref{ka}.

The eigenvalues $\lambda^{(i)}$ of the $\chi-$matrix can be calculated in this case, indeed
\begin{align}
    \lambda^{(1,2)}=\frac{1+|h|^2+|k|^2}{2}\pm\sqrt{\left(\frac{1+|h|^2+|k|^2}{2}\right)^2-|k|^2},
\end{align}
and
\begin{align}
    \lambda^{(3,4)}=\frac{1-|h|^2-|k|^2}{2}\pm\sqrt{\left(\frac{1-|h|^2-|k|^2}{2}\right)^2+|k|^2}.
\end{align}
Note that $\lambda^{(4)}<0$ if $k\neq 0$. Hence $\bm\chi_{ab}$ is never positive semi-definite unless either
\begin{align}
    \abs{\bm r(t_1)\times\bm r(t_0)}=0,
\end{align}
when the $\bm r(t_1)$ is parallel to  $\bm r(t_0)$, or 
\begin{align}
    \braket{\hat{O}(t_1)\hat{O}(t_0)}_{\hat{\rho}_{\textsc{e}}}=0,
\end{align}
when the correlations on the environment vanish. Therefore, if we assume that the correlator does not vanish for any $t_1,t_2$, the process is not CP-divisible in general.

Note that if $\bm r(t_1)\times\bm r(t_0)=0$, the channel \eqref{CPTP2d} is just a pure phase-damping channel, which can be solved for an arbitrary number of interactions as explained in section \ref{dephasing}.  However, for completeness, let us consider such special case  here as well. Setting \mbox{$\abs{\bm r(t_1)\times\bm r(t_0)}=0$}, the eigenvalue $\lambda^{(4)}$ is given by
\begin{align}
  \lambda^{(4)}= \Big|1-|h|^2\Big|\frac{\text{sgn}(1-|h|^2)-1}{2},
\end{align}
which is negative whenever $|h|>1$. In contrast with the general case, complete positivity is not always lost but it depends on the strength of the correlations relative to the self correlations in the environment.

Since CP-divisibility implies P-divisibility, hence we now focus on determining in which cases the process is not even P-divisible, i.e., when the map \eqref{intermap} is not positive.  Note that a necessary and sufficient condition for a trace-preserving map (such as \eqref{intermap}) to be positive can be written in terms of its action over the Bloch vector in the initial state. Namely,  $\Theta_{t_1,t_0}$ is positive if and only if 
\begin{align}
    \abs{\bm \Theta_{t_1,t_0}(\bm u)}\leq 1
\end{align}
$\forall \bm u$. Note that the norm of the outcome of the map is a convex function over the Bloch sphere, therefore its image is contained in the image of the map acting over the set  of pure states, i.e. the set $\abs{\bm u}=1$. Thus, the map is positive if and only if the image of the surface $\abs{\bm u}=1$ is contained in the Bloch sphere. Similarly to the channel \eqref{CPTP2dBloch}, the map can be written as an affine transformation in the basis \eqref{vecbasis}. Note that in order to obtain this affine transformation we can `recycle' a calculation we already made. In particular we only have to set $g=1$ by hand in equation \eqref{Chunga}. Indeed,
\begin{align}
    \bm \Theta_{t_1,t_0}(\bm u)=A\bm u+\bm b,
\end{align}
where
\begin{align}
A=
    \begin{bmatrix}
     1 &  0& 0 \\
    2\Re(h k^*)      & |h|^2-|k|^2 & 0 \\
 0 & 0 &|h|^2 +|k|^2
\end{bmatrix}
\end{align}
and
\begin{align}
\bm{b}=
    \begin{bmatrix}
     0 \\
     0 \\
  2\Im(h k^*)
\end{bmatrix}.
\end{align}
Now take $\bm u=\bm r(t_1)$. Its image under the channel is
\begin{align}
 &\nn\bm \Theta_{t_1,t_0}(\bm r(t_1))=\bm r(t_1)+ 2\Im(h k^*) \frac{\bm r(t_1)\times\bm r(t_0)}{\abs{\bm r(t_1)\times\bm r(t_0)}}+\\
 &2\Re(h k^*) \frac{\left(\bm{r}({t_0})-(\bm{r}({t_1})\cdot\bm{r}({t_0}))\bm{r}({t_1})\right)}{\abs{\bm r(t_1)\times\bm r(t_0)}}.
\end{align}
But then
\begin{align}
     \abs{\bm \Theta_{t_1,t_0}(\bm r(t_1))}^2=1+4|h k|^2\geq1,
\end{align}
where the equality is saturated if and only if either $k=0$ or $h=0$. But $h\neq0$, as it can be easily seen from its expression in \eqref{hache}. Hence, the map is not positive unless $k=0$, which is also a condition for complete divisibility. Now, we address the general case when $k=0$, where the map may be written as
\begin{align}
    \bm \Theta_{t_1,t_0}(\bm u)=   
    \begin{bmatrix}
     1 &  0& 0 \\
    0    & |h|^2 & 0 \\
    0 & 0 &|h|^2 
\end{bmatrix}\bm u.
\end{align}
Since in this case the matrix is diagonal in the basis given by \eqref{basis}, it easily follows that the channel is positive if and only if $|h|\leq1$. Since this was also true for complete divisibility, we have shown that divisibility and complete divisibility are equivalent for the process.

Now we turn our attention to the pure dephasing scenario. Recall that, for this case, the channel is just
\begin{align}
  \nn   \mathcal{E}_{t_n\dots t_0}[\hat{\rho}_{\textsc{q}}^0]&=\frac{1}{2}(\text{Id}+\mathcal{U})[\hat{\rho}_{\textsc{q}}^0]\\
  & + \frac{\Re \gamma}{2}(\text{Id}-\mathcal{U})[\hat{\rho}_{\textsc{q}}^0]
  +\frac{\ii\Im \gamma}{2}\left[\bm{r}\!\cdot\!\bm{\hat{\sigma}},\hat{\rho}\right],
\end{align}
where, again, $\mathcal{U}[\hat\rho]=(\bm{r}\!\cdot\!\bm{\hat{\sigma}})\hat\rho(\bm{r}\!\cdot\!\bm{\hat{\sigma}})$ and $\gamma\equiv\gamma_{+-}$ is the dephasing parameter defined in equation \eqref{gammadeph}. 

This channel always admits an inverse as far as $\gamma\neq0$, indeed
\begin{align}
  \nn   \mathcal{E}_{t_n\dots t_0}^{-1}[\hat{\rho}_{\textsc{q}}^0]&=\frac{1}{2}(\text{Id}+\mathcal{U})[\hat{\rho}_{\textsc{q}}^0]\\
  & + \frac{\Re \gamma}{2\abs{\gamma}^{2}}(\text{Id}-\mathcal{U})[\hat{\rho}_{\textsc{q}}^0]
  -\frac{\ii\Im \gamma}{2\abs{\gamma}^{2}}\left[\bm{r}\!\cdot\!\bm{\hat{\sigma}},\hat{\rho}\right].
\end{align}
In this case, a different number of interactions is described by the same channel with a different parameter $\gamma$. Let us define $\gamma_n$ the dephasing parameter corresponding to $n$ interactions. Thus, if we are to compare two pure dephasing processes associated with different numbers of interactions, say, one with $n$ interactions and another with $m>n$, the transition map is given by
\begin{align}
    \nn&\Theta_{t_m\dots t_0}=\mathcal{E}_{t_m\dots t_0}\circ\mathcal{E}^{-1}_{t_n\dots t_0}\\
    &\nn=\frac{1}{2}(\text{Id}+\mathcal{U})[\hat{\rho}_{\textsc{q}}^0]\\
  & + \frac{\Re (\gamma_m\gamma_n^*)}{2\abs{\gamma_n}^{2}}(\text{Id}-\mathcal{U})[\hat{\rho}_{\textsc{q}}^0]
  +\frac{\ii\Im (\gamma_m\gamma_n^*)}{2\abs{\gamma_n}^{2}}\left[\bm{r}\!\cdot\!\bm{\hat{\sigma}},\hat{\rho}\right].
\end{align}
In this case, the $\chi-$matrix representation of the transition map is straightforward to compute. We can choose an orthonormal basis consisting of 
\begin{align}\label{basispd}
   \nn& \hat B_0=\frac{\openone}{\sqrt{2}}, & \hat B_1=\frac{\bm{r}\cdot\bm{\hat{\sigma}}}{\sqrt{2}},&
\end{align}
plus some orthonormal completion.

In such basis, the $\chi-$matrix representation of $\Theta_{t_m\dots t_0}$ takes the form
\begin{align}
\bm{\chi}=
    \begin{bmatrix}
     1+\Re\frac{\gamma_m}{\gamma_n}  &  \ii\Im\frac{\gamma_m}{\gamma_n} & 0 &  0 \\
    -\ii\Im\frac{\gamma_m}{\gamma_n}     & 1-\Re\frac{\gamma_m}{\gamma_n} & 0 & 0 \\
   0 &0&  0  &0\\
 0   & 0 & 0 &  0
\end{bmatrix}.
\end{align}
Its eigenvalues are straightforward to compute, 
\begin{align}
    \lambda^{1,2}=1\pm\frac{\abs{\gamma_m}}{\abs{\gamma_n}},\qquad
\lambda^{3,4}=0,
\end{align}
and from them we know that the transition map is completely positive if and only if $\abs{\gamma_m}\leq\abs{\gamma_n}$.

This is not surprising, since the condition $\abs{\gamma_m}\leq\abs{\gamma_n}$ just states that the non-diagonal terms  of the spin's state (in the basis where $\bm{r}\cdot\bm{\hat{\sigma}}$ is diagonal, see \eqref{commonsd}) has to decrease monotonically at each step. Recall that the pure dephasing scenario is formally equivalent to the case of a single interaction described in section \ref{single}. It was shown there that the purity given by equation \eqref{PurityFirst}, and thus the entanglement entropy, is a monotonic function of the parameter $\gamma$. We conclude that divisibility implies that the purity is a monotonic function of the number of interactions, thus no information can be recovered from the environment as the number of interactions increases.

\section{conclusions and outlook}\label{conclusion}

In this paper we have made use of algebraic properties of bosonic systems to achieve  non-perturbative results for the dynamics of a two-level system interacting with a bosonic reservoir. In order to achieve non-perturbative results without any approximations we have considered a sequence of fast and intense couplings modeled with delta distributions in the time domain. We have proved that for Gaussian states of the reservoir these couplings induce a CPTP map that acts over the spin in a way that mimics a collisional model. 

First, we have completely characterized the  dynamics of the two-level system for a single fast interaction, showing that the first moment of the bosonic operator $\hat O$ induces a unitary channel over the spin whereas the second moment induces a phase damping. We have argued how the non-commutative character of the environment observable at different times is irrelevant for the dynamics. We have also shown that through a Dirac-delta interaction the spin gets entangled with the environment, and its entanglement entropy can be evaluated non-perturbatively as a function of the intensity of the fluctuations in the bosonic state.

Next, we have written the quantum channel describing the dynamics of the two-level system as the partial trace of a sequence of controlled displacements over the bosonic system, and we have given a closed expression for the channel. More concretely, we have provided a way to calculate the so-called $\chi$ matrix representation of the channel. Moreover, we have studied the role of environmental time correlations in the factorization of the channel. Namely, we showed that in the limit where the time correlations of the environment vanish at all times the channel can be thought of as a ``bombardment''  of ancillary systems (See e.g., \cite{PhysRevA.97.052120}). In that limit the channel becomes just an iterated application of the single-interaction channel.  We have also carried out a brief study of the fixed points of such bombardments.

Then, we have fully characterized the channel with two fast interactions, showing that it can be explicitly written in terms of the real and imaginary parts of the time correlations of the environment. We have shown that, if the correlators have a non-vanishing imaginary part, the channel is not unital, thus the spin's purity can increase for some mixed initial states. In other words the spin can be purified by repeatedly interacting in short bursts with an environment observable with a non-vanishing different-time commutator.

We have also studied scenarios in which an arbitrary number of interactions induces a pure dephasing channel. Namely, we have shown that if the the spin's degrees of freedom in the interaction Hamiltonian commute at the specific times of interaction, then the channel is a pure dephasing channel. We have argued that this situation may be engineered  by sinchronizing the interactions with the free dynamics of the spin. In addition, we have analyzed the role of time ordering in these scenarios.

Finally, we have studied non-Markovian effects in the dynamics induced in the spin. More concretely, we have analyzed divisibility, i.e. the complete positive character of the transition maps between two arbitrary interactions. Also, we have studied divisibility for an arbitrary number of synchronized interactions in pure dephasing channels. For the two-interactions channel, we have shown that complete divisibility is equivalent to divisibility, and that the channel is not divisible if the imaginary part of the time correlations of the environment is non-zero. If the relation between backflow of information and divisibility is to be taken seriously, this means that with a correlated environment there exist states of the spin such that information that is lost to the environment can return to the spin. We have also shown that in pure dephasing situations the lack of divisibility is equivalent to the monotonicity of the purity of the spin's state as a function of the number of interactions. 

The study of these fast-interactions from the perspective of open quantum systems  may reveal helpful in relativistic quantum information. Relativistic quantum information is concerned with the dynamics of quantum information in relativistic scenarios, and it is natural, and common, to frame its results within the formalism of quantum field theoy (QFT). Our model is well-suited for applications in the so-called particle detector models, a family of models in which a detector system interacts locally with a quantum field, usually chosen to be bosonic for simplicity \cite{Hawking:1979ig,PhysRevD.97.105026,Crispino}.

Falsifiable predictions in quantum field theory correspond, in many the cases, to scattering processes in which there is no notion of continuous evolution that describes the state of the system at a intermediate time. This can be problematic for two reasons. First, QFT is plagued with ultraviolet divergences that are only tractable if the observables of the theory are smooth functionals over space and time \cite{MR0493420,Louko_2006,Satz_2007}. Second, projective measurements are forbidden in quantum field theory, since they violate causality \cite{Sorkin:aa}, thus it is conceptually troublesome to address the statistics of the system while the interaction is still switched. 

Indeed, Markovianity in QFT has raised interest in the community of relativistic quantum information in the past (See, among many others, \cite{Moustos:2017aa,Sokolov:2018aa,Benatti:2004aa,Lin:2007aa}). In these works  master equations were used to study thermalization of accelerated detectors when interacting with the vacuum of a free scalar field, and it was studied also  the role of memory effects caused by more general trajectories. The study of Markovianity from the perspective of comparing scattering processes, as it was presented in this paper, may prove helpful in the formalization of an approach to non-Markovian effects in QFT, at least from the perspective of particle detector models.

Note that, despite the initial motivation of applying our results to particle detector models, we have been deliberately ambiguous in the physical interpretation of the bosonic degrees of freedom. Indeed, we expect our results to apply in a wide set of physical scenarios.  We believe that in other set-ups, e.g. the light-matter interaction in quantum optics, our results may provide a good theoretical framework for designing interactions that induce a desirable set of quantum channels by controlling the coupling of qubits with an environment or the time correlations of the environment itself. 

Regarding future work, we expect to apply the results of this paper to the study of particle detectors in quantum field theory, in particular to  relativistic classical and quantum communication (see, among many others, \cite{ClicheKempfD,Jonsson1,Jonsson2,Jonsson3,Jonsson4,Bla,Katja,Landulfo,Shockwaves,Petar}) and its interplay with backflow of information between the detectors and the field. An interesting avenue to explore is the limits in which we could use these results to approximate smooth switching functions (many short kicks that approximate a continuous interaction), which could be efficiently evaluated numerically. Such results should be achievable as can be argued from the proofs of the theorems concerning integration of dynamical systems in infinite dimensions \cite{KATO_1953}. Obtaining this result can be helpful for the non-perturbative treatment of the spin-boson model in general and for calculations regarding detectors in curved spacetimes.

\acknowledgments

  The authors thank Angel Rivas for his help reviewing the extensive literature of open quantum systems. EMM acknowledges support through the Discovery Grant Program of the Natural Sciences and Engineering Research Council of Canada (NSERC). EMM also acknowledges support of his Ontario Early Researcher award.

\appendix

\section{$\chi$ matrix representation}\label{chiap}
In this appendix we briefly introduce the $\chi$ matrix representation \cite{nambu2005matrix} and particularize a formula for the representation of the channel in Eq. \eqref{CPTP2}. In general the complete positive property channel can be encoded in this representation through simpler properties of linear maps between finite dimensional systems, and in this work we make use of it to characterize divisibility.  
In this representation, the action of a the linear map the is written as the sum
\begin{align}\label{channelando}
   \mathcal{E}_{t_N...t_0}[\hat{\rho}_{\textsc{q}}^0]= \sum_{a b}\bm\chi_{ab} \hat B_a\hat{\rho}_{\textsc{q}}^0\hat B_b^\dagger,
\end{align}
where the $\{\hat B_i\}$ is a  basis of operators that are orthonormal with respect to the inner product 
\begin{align}\label{inner}
    \langle\hat A,\hat B\rangle=\tr(\hat B^\dagger \hat A),
\end{align}
and $\bm\chi_{ab}$ is the $\chi-$matrix. The $\chi-$matrix in qubit channels is a four by four matrix, therefore the indices $a$ and $b$ run from $0$ to $3$.

For our particular channel, the entries of $\bm\chi_{ab}$ can be computed from \eqref{CPTP2} and  \eqref{channelando} as
\begin{align}\label{chi}
    \nn\bm\chi_{ab}=&\sum_{\{s_i,s'_i=1,-1\}}\!\!\!\!\!\!\gamma_{_{\{s_i,s'_i\}}}\\
    &\times\text{tr}\left(\hat B_a^\dagger\hat{P}^{s_N}({t}_N)...\hat{P}^{s_0}({t}_0)\right)\\
    & \nn\times\text{tr}\left(\hat{P}^{s'_0}({t}_0)...\hat{P}^{s'_N}({t}_N)\hat B_b\right),
\end{align}
where we have expanded the products of projectors in the orthonormal basis,
\begin{align}
    \hat{P}^{s_N}(t_N)...\hat{P}^{s_0}(t_0)=\sum_{a=0}^3\tr(\hat B_a^\dagger \hat{P}^{s_N}(t_N)...\hat{P}^{s_0}(t_0))\hat B_a.
\end{align}

\section{Formal derivation of the collision-like unitary evolution}\label{deltasproof}
In this appendix we will prove that the train of delta switchings indeed generates a time ordered product of unitary operators.  

Let us consider that the distributional limit to the delta interaction is taken in the following way
\begin{align}\label{trensuave}
    \chi_{\delta t} (t)&=\frac{1}{{\delta t}}\sum_{k=0}^N \xi\left(\frac{t-{t}_k}{{\delta t}}\right) & {t}_N>&...>{t}_0 
    \end{align}
where $\xi$ is a positive, symmetric and normalized (in the sense of $\int \xi(t) \d t=1$) function of time. When  $\delta{t} \to 0$ \eqref{trensuave} tends,  in the distributional sense, to the train of delta interactions described by \eqref{deltas}.

The objective of this appendix is to show that he evolution generated by such a switching function acquires the following form when $\delta{t} \to 0$:
\begin{align}\label{limit}
  \hat{U}_{I}=  \lim_{{\delta t}\to0} \hat{U}_{\delta t}=&e^{-\ii \hat{H}_{I}( {t}_N)}...e^{-\ii \hat{H}_I( {t}_0)} & {t}_N>&...>{t}_0.
\end{align}

To do this, we first express the unitary evolution in terms of its Dyson series, that is,
\begin{align}\label{Dysonhamiltonian}
&\nn\hat{U} _{\delta t}=\\
&\sum_{n=0}^{\infty}\frac{(-\ii)^n}{n!}\!\!\int...\int\d^n t\, \mathcal{T}\left[\hat{H}(t_1)...\hat{H}(t_n)\right]\chi_{\delta t}(t_1)...\chi_{{\delta t}}(t_n).   
\end{align}
Note that we could extract the functions $\chi(t)$ from the time-ordered product, since they are not operators.
Next, we consider the integral
\begin{align}
&\nn I_n=\frac{1}{{\delta t}^n}\!\!\!\!\int...\int\d^n t\; \mathcal{T}\left[\hat{H}(t_1)...\hat{H}(t_n)\right]\;\chi_{\delta t}(t_1)...\chi_{\delta t}(t_n) \\
&=\frac{1}{{\delta t}^n} \int\!...\!\int\d^n t\, \mathcal{T}\left[\hat{H}(t_1)...\hat{H}(t_n)\right]\!\prod_{i=1}^{n} \sum_{k=0}^N \xi\left(\frac{t_i-{t}_k}{{\delta t}}\right).
\end{align}
Note that
\begin{align}
 \prod_{i=1}^{n} \sum_{k=0}^N \xi\left(\frac{t_i-{t}_k}{{\delta t}}\right)=
 \sum_{k_1=0}^N...\sum_{k_n=0}^N \prod_{i=1}^{n}\xi\left(\frac{t_i-{t}_{k_i}}{{\delta t}}\right).
\end{align}
Therefore, the integral becomes the sum
\begin{align}
\nn I_n =\frac{1}{{\delta t}^n}\sum_{k_1=0}^N...\sum_{k_n=0}^N \int...&\int\d^n t\; \mathcal{T}\left[\hat{H}(t_1)...\hat{H}(t_n)\right]\\
&\times \prod_{i=1}^{n}\xi\left(\frac{t_i-k_i{t}}{{\delta t}}\right). 
\end{align}
Now we perform the following changes of variables in each of the terms:
\begin{align}
    \bar{t}_i=\frac{t_i-{t}_{k_i}}{{\delta t}}.
\end{align}
Under these changes of variables the integral becomes
\begin{align}
\nn I_n&=\sum_{k_1=0}^N...\sum_{k_n=0}^N \int...\int\d^n t\;\\ &\times\mathcal{T} \left[\hat{H}({\delta t}\bar t_1 +{t}_{k_1})...\hat{H}({\delta t}\bar t_n +{t}_{k_n})\right]\prod_{i=1}^{n}\xi\left(\bar{t}_i\right). 
\end{align}
 Assuming that we can take the limit ${\delta t}\to0$ inside the integral sign, we obtain the following result:
\begin{align}\label{mierdimite}
\lim_{{\delta t}\to0}I_n=\sum_{k_1=0}^N...\sum_{k_n=0}^N  \mathcal{T}\left[ \hat{H}({t}_{k_1})...\hat{H}({t}_{k_n})\right]. 
\end{align}
As the terms are time-ordered, we can express the sum as 
\begin{align}\label{multinomhamiltonian}
\nn\lim_{{\delta t}\to0}I_n=\sum_{k_1=0}^N...\sum_{k_n=0}^N  \mathcal{T}\left[ \hat{H}({t}_{k_1})...\hat{H}({t}_{k_n})\right]\\
=\sum_{\{\sum s_i=n\}}\binom{n}{s_0...s_n} \hat{H}^{s_N}({t}_N)...\hat{H}^{s_0}({t}_0).
\end{align}
Next, we introduce the formula \eqref{multinomhamiltonian} in expression \eqref{Dysonhamiltonian}, thus obtaining
\begin{align}
U_{\delta}=\sum_{n=0}^{\infty}\frac{(-\ii)^n}{n!}   \sum_{\{\sum s_i=n\}}\binom{n}{s_0...s_n} \hat{H}^{s_N}({t}_N)...\hat{H}^{s_0}({t}_0).
\end{align}
A last rearrangement of the terms leads to the desired result, that is
\begin{align}
\nn U_{\delta}&=\sum_{n=0}^{\infty}\frac{(-\ii)^n}{n!}   \sum_{\{\sum s_i=n\}}\binom{n}{s_0...s_n} \hat{H}^{s_N}({t}_N)...\hat{H}^{s_0}({t}_0)\\
&=\prod_{k=0}^N \sum_{s=0}^{\infty}\frac{(-\ii)^s}{s!}\hat{H}^{s}({t}_{k})=e^{-\ii \hat{H}( {t}_N)}...e^{-\ii \hat{H}( {t}_0)}.
\end{align}

This proof is valid in general only for bounded operators. However, in this paper only Gaussian states, such that for instance coherent, KMS and ground states, are considered. It can be shown that all these states are analytic vectors for a CCR representation \cite{MR0493420}, thereby the results in terms of power series are still valid in the strong convergence sense.

\section{Unitality and entropy gain}\label{entropygain}
In this appendix we briefly discuss the relation between the unitality of a quantum channel and the lower bound on the entropy gain under its action. For a general, rigorous analysis of this implication, we refer the reader to \cite{Lindblad_1973}.

It is known that the relative entropy is monotonic under CPTP channels. Indeed, the relative entropy between two Hilbert-Schmidt operators $\rho$ and $\sigma$, defined as
\begin{align}
    S(\rho||\sigma)=\text{tr}[\rho \log \rho]-\text{tr}[\rho \log \sigma],
\end{align}
fulfills
\begin{align}
    S(\mathcal{E}[\rho]||\mathcal{E}[\sigma])\leq S(\rho||\sigma).
\end{align}
Therefore, if we fix $\sigma=I$ we get
\begin{align}
    \nn S(\mathcal{E}[\rho]||\mathcal{E}[I])&=-\text{tr}[\mathcal{E}[\rho] \log \mathcal{E}[I]]-S(\mathcal{E}[\rho])\\
    &\leq S(\rho||I)=-S(\rho),
\end{align}
since $\log I=0$. Finally, we rearrange the inequality, thus obtaining
\begin{align}\label{unitalinc}
    S(\mathcal{E}[\rho])-S(\rho)\geq\text{tr}[\mathcal{E}[\rho] \log \mathcal{E}[I]].
\end{align}
For a unital channel the right hand side of equation \eqref{unitalinc} vanishes, therefore
\begin{align}
    S(\mathcal{E}[\rho])\geq S(\rho).
\end{align}

\section{Pure dephasing in general scenarios}\label{puredephasing}
This appendix is devoted to studying general pure dephasing dynamics, and to show that the irrelevance of the time ordering in the pure dephasing dynamics we found in the main text is not present for arbitrary systems coupling to bosonic baths, but rather it is inherent to the two-level nature of the spin.

Let us consider a more general interaction Hamiltonian
\begin{equation}
    \hat{H}(t)=\hat{m}(t)\otimes \hat{O}(t)\chi(t),
\end{equation}
where this time $\hat{m}(t)$ is a general finite-dimensional self-adjoint operator.

Generally speaking, dependence on the time-order will arise in scenarios where the eigenvalues of the system's coupling observable $\hat{m}(t)$ do not lie on the unit circle of the complex plane, i.e. when the coupling observable is not unitary. Since the spin-boson coupling Hamiltonian has to be self-adjoint, independence of time-ordering can be achieved only when the system couples to its environment through unitary, self-adjoint operators e.g. the constituents of the Pauli group.

We observe that the fact that time ordering only implements a ``controlled" phase respect to the different eigenstates of the system's observable does not come from the particularities of the switching function. Instead, it follows from the constraint on$\hat m(t)$ to commute with itself at all times (for the pure dephasing channel). Indeed, consider the global unitary evolution of system+environment, but with a general system's observable $\hat{m}(t)$:
\begin{align}
  \hat{U}=  \mathcal{T}e^{-\ii\int\d t \hat{m}(t)\otimes \hat{O}(t)\chi(t)}.
\end{align}
Now, consider that $[\hat{m}(t),\hat{m}(t')]=0$ for all $t,t'\in \text{supp}[\chi]$. Then, we can extract the effect of time-ordering with the  Magnus expansion \cite{magnus1954exponential,Blanes_1998,B_tkai_2011}. Namely, given a one-parameter family of self-adjoint operators $\hat{A}(t)$, it holds that 
\begin{align}\label{magnus}
  \mathcal{T}e^{-\ii\int\d t \hat A(t)}= e^{-\ii\int\d t \hat A(t)-\int\d t \int^{t}\d t' [\hat{A}(t),\hat{A}(t')]+\dots},
\end{align}
where the dots denote terms that involve at least two commutators. Since $[\hat{m}(t),\hat{m}(t')]=0$, the commutator of the whole interaction Hamiltonian with itself at different times is proportional to $\hat{m}(t)\hat{m}(t')$:
\begin{align}
     \nn&[\hat{m}(t)\otimes \hat{O}(t)\chi(t), \hat{m}(t')\otimes \hat{O}(t')\chi(t')]\\
     &= \chi(t)\chi(t')\hat{m}(t)\hat{m}(t')\otimes [\hat{O}(t),\hat{O}(t')]\\
    & =\chi(t)\chi(t')\mathcal{C}(t,t')\hat{m}(t)\hat{m}(t')\otimes\openone_{\textsc{e}} 
\end{align}
and the Magnus expansion only involves the first two terms in the exponent of \eqref{magnus}. Explicitly
\begin{align}
  &\nn\hat{U}= e^{-\ii\int\d t \hat{m}(t)\otimes \hat{O}(t)\chi(t)}\\
  &\times e^{-\int\d t \int^{t}\d t'\chi(t)\chi(t')\hat{m}(t)\hat{m}(t')\mathcal{C}(t,t') }.
\end{align}
Finally, since $[\hat{m}(t),\hat{m}(t')]=0$, $\hat{m}(t)$ admits a time independent spectral decomposition of the form
\begin{align}
    \hat{m}(t)=\sum_{i}\lambda_i(t) \hat P_i,
\end{align}
and thus the joint unitary evolution is given by
\begin{align}
  &\nn\hat{U}= \sum_i  e^{-\int\d t \int^{t}\d t'\chi(t)\chi(t')\lambda_i(t)\lambda_i(t')\mathcal{C}(t,t') }\\
  &\times\hat P_i\otimes e^{-\ii\int\d t \lambda_i(t) \hat{O}(t)\chi(t)}.
\end{align}

\bibliography{refs.bib}

\end{document}